\documentclass[]{aa}  
\usepackage{graphicx}
\usepackage{txfonts}
\usepackage{natbib}
\bibpunct{(}{)}{;}{a}{}{,} 

\begin{document}
   \title{Modelling magnetic flux emergence in the solar convection zone}
   \titlerunning{Modelling flux emergence}
   \author{P.J. Bushby	
          \inst{1}
          \and
          V. Archontis\inst{2}
}
   \institute{School of Mathematics and Statistics, Newcastle University,
              Newcastle Upon Tyne, NE1 7RU, UK\\
              \email{paul.bushby@ncl.ac.uk}
         \and
             School of Mathematics and Statistics, University of St. Andrews, North Haugh,
             St. Andrews, Fife, KY16 9SS, UK\\
             \email{vasilis@mcs.st-and.ac.uk}
 }     
       
 
  \abstract
{Bipolar magnetic regions are formed when loops of magnetic flux
  emerge at the solar photosphere. Magnetic buoyancy plays a crucial
  role in this flux emergence process, particularly at larger scales. However it is not yet clear to
  what extent the local convective motions influence the evolution of
  rising loops of magnetic flux.} 
{Our aim is to investigate the flux emergence process in a simulation
  of granular convection. In particular we aim to determine the circumstances under which magnetic buoyancy enhances the flux emergence rate (which is otherwise driven solely by the convective upflows).} 
{We use three-dimensional numerical simulations, solving the equations
  of compressible magnetohydrodynamics in a horizontally-periodic
  Cartesian domain. A horizontal magnetic flux tube is inserted into
  fully developed hydrodynamic convection. We systematically vary the
  initial field strength, the tube thickness, the initial entropy
  distribution along the tube axis and the magnetic Reynolds number.} 
{Focusing upon the low magnetic Prandtl number regime ($Pm<1$) at moderate magnetic Reynolds number, we
  find that the flux tube is always susceptible to convective disruption to some extent. However, stronger flux tubes tend to maintain their structure more effectively than weaker ones. Magnetic buoyancy does enhance the flux emergence rates in the strongest initial field cases, and this enhancement becomes more pronounced when we increase the width of the flux tube. This is also the case at higher magnetic Reynolds numbers, although the flux emergence rates are generally lower in these less dissipative simulations because the convective disruption of the flux tube is much more effective in these cases. These simulations seem to be relatively insensitive to the precise choice of initial conditions: for a given flow, the evolution of the flux tube is determined primarily by the initial magnetic field distribution and the magnetic Reynolds number.} 
{}
\keywords{Convection -- Magnetohydrodynamics (MHD) -- Sun: granulation -- Sun: interior}

\maketitle
%

\section{Introduction}
It is generally believed that most of the large-scale toroidal
magnetic field in the solar interior is stored in the
stably-stratified layer just below the base of the convection zone,
where the toroidal field is amplified by the effects of differential rotation
\citep[see, for example,][]{OSSEN03}. Through the action of magnetic
buoyancy \citep[][]{PARK55}, segments of this toroidal flux rise into
the convection zone, eventually emerging to form bipolar active regions at the solar surface \citep[][]{ZWAAN85}. Several numerical simulations of this
flux emergence process have focused upon the interactions between
convection and buoyant magnetic flux tubes
\citep[][]{DORCH01,FAN03,ABBETT04,STEIN11}, whilst  others have
focused upon the subsequent small-scale and large-scale evolution of
the emerging field, as it  rises into the photosphere/chromosphere
\citep[e.g.][]{CSM07, TMI09} and into the lower corona
\citep[e.g.][]{MAG01,FAN01,ARC04,SYK08}. There have also been a number
of recent reviews of flux emergence dynamics
\citep[e.g.][]{HOOD11,ARC12}. 

\par Since it is not yet possible to model the whole flux emergence process (from the base of the solar convection zone into the solar atmosphere) in a self-consistent way, it is necessary to consider local models with idealised initial conditions. Usually, the initial magnetic field distribution corresponds to a thin horizontal flux tube. This flux tube is often twisted, although the components of the magnetic field that are perpendicular to the tube axis are usually much weaker than the axial component. As a result, the dominant component of the Lorentz force corresponds to the radial gradient of the magnetic pressure. In calculations of this type, the local gas pressure is usually modified so that this initial flux tube is in approximate (total) pressure balance with its non-magnetic surroundings. The tube is then made buoyant by specifying a
particular entropy distribution (or, equivalently, a density deficit) along its axis. In some models, the imposed specific entropy distribution varies along the axis of the tube, with the result that only one part of the tube is buoyant. This differential buoyancy  leads to the development of $\Omega$-like loops of magnetic flux. When convective motions are present, it is not necessary to impose a variable entropy distribution along the tube in order to obtain $\Omega$-like loops. Convective upflows (and downflows) naturally lead to the formation of loop-like structures, provided that the magnetic Reynolds number is large enough to ensure that the magnetic field lines are (at least partially) advected with the flow.  

\par In the absence of convective motions, there is nothing to disrupt
the buoyant rise of a magnetic flux tube. The situation is clearly
more complicated when the tube rises through a convective layer. In a
previous study, \citet{FAN03} used the anelastic approximation to
model the interaction between uniformly buoyant magnetic flux tubes
and convection in the deep layers of the solar convection
zone. Varying the initial field strength in their model, they found
that weak fields tended to be disrupted by the convective motions. In
fact the peak field strength within the tube had to be significantly
greater than the equipartition value (at which the field is in energy
balance with the surrounding motions) before magnetic buoyancy could
play a dominant role in the evolution of the flux tube. Similar
conclusions were reached by \citet{ABBETT04} in their related
study. Although we do not go into more details here, it is worth
noting one further aspect of these anelastic models. Defining the
magnetic Prandtl number, $Pm$, to be the ratio of the magnetic
Reynolds number to the (fluid) Reynolds number, \citet{FAN03} focused
exclusively upon the $Pm>1$ regime, at relatively low Reynolds
numbers. The simulations of \citet{ABBETT04} were at higher Reynolds
number, but most were still in this $Pm>1$ regime. In actual fact, we
would expect $Pm\ll 1$ throughout the solar convection zone
\citep[see, e.g.][]{OSSEN03}. However, the low magnetic Prandtl number
regime is numerically extremely challenging, which presumably explains
why these studies did not focus upon this region of parameter space. 

\par Models of flux emergence through fully compressible convection
have also been carried out. It is possible to simulate this process on
scales that are comparable to that of an active region
\citep[][]{CRTS10,STEIN11}. However, it is not feasible to carry out
parametric surveys in calculations of this size, so most previous
studies have focused upon flux emergence through convective layers
with a relatively small number of granular cells. The key finding of
the compressible model of \citet{DORCH01} is that convective
disruption removes magnetic flux from the tube as it rises, thus
reducing the quantity of flux that emerges at the surface layers. In
some sense, therefore, convective motions reduce the efficiency of the
flux emergence process. However, it is not clear to what extent this
effect is parameter-dependent. \citet{CSM07} considered a model of
compressible convection that included the effects of radiative
transfer. At some instant in time, they introduced a long,
uniformly-buoyant magnetic flux tube into the lower part of the
domain, modifying the local gas pressure and the local velocity field
so that the tube was (at least instantaneously) in equilibrium with
its non-magnetic surroundings. Under the action of the vertical
convective motions, they found that weak flux tubes tend to develop a
sea-serpent-like form. However, as in the anelastic studies, magnetic
buoyancy dominates the evolution of the tube only
when the peak magnetic field strength exceeds some threshold value. At
the surface these emerging fields can be strong enough to modify the
granulation pattern. In their larger scale simulations,
\citet{STEIN11} adopted a different approach, introducing untwisted,
horizontal magnetic flux through the lower boundary of their
convective domain, ensuring that it had the same entropy as the
convective upflows. Like \citet{CSM07}, they found that weak fields
were susceptible to convective disruption. Fields that were strong
enough to rise to the surface through the action of magnetic buoyancy
tended to modify the surrounding flow, producing unrealistically hot,
large granules.   

\par Motivated by these previous studies, the aim of this paper is to
investigate the parametric dependence of simulations of flux emergence
across a convectively-unstable compressible layer in the ``low"
magnetic Prandtl number regime. Most of the compressible calculations
that are described above utilise artificial viscosities to stabilise
the numerical scheme, so it is difficult to define a magnetic Prandtl
number in these cases although, in practice, it is likely that $Pm
\approx 1$ in these calculations. Therefore, this will be the first
time that the $Pm <1$ regime has been studied systematically in this
context. Given that {\it any} initial conditions in local models of
this type are likely to be highly idealised, it is important to
determine the extent to which the evolution of the system is sensitive
to the precise choices that are made. We therefore consider several different
initial configurations, varying the peak field strength, the width
(and twist) of the tube, in addition to varying the initial entropy
distribution within the flux tube. By carrying out this systematic
survey we hope to identify the key features of the initial configuration that determine whether or not magnetic buoyancy contributes towards the evolution of the tube. The paper is
structured as follows. In the next section, we describe the governing
equations, the model parameters and the choice of initial conditions
under consideration. In Section 3, we present our numerical
results. In the last section, we summarise our findings and discuss
some of their implications for flux emergence in the solar convection zone.  

\section{Model setup}

\subsection{Governing equations}
We consider the dynamics of a plane layer of compressible,
electrically-conducting fluid that is heated from below. The fluid is
characterised by various parameters (all of which we assume to be
constant in this idealised model): the thermal conductivity $K$, the
shear viscosity $\mu$, the magnetic diffusivity $\eta$, the magnetic
permeability $\mu_0$, and the specific heat capacities, $c_V$ and
$c_P$. The gas constant, $R_*$, is defined by $R_*=c_P-c_V$. Choosing
a Cartesian coordinate system in which the $z$-axis points vertically downwards
(parallel to the constant gravitational acceleration
$g\mathbf{\hat{z}}$), this fluid occupies the region $0 \le x\le 8d$,
$0\le y \le 8d$ and $0\le z\le d$, where $d$ is the depth of the
domain. Periodic boundary conditions are imposed in the horizontal
directions. The upper and lower boundaries are held at fixed
temperature, so that $T=T_0$ at $z=0$ and $T=T_0+\Delta T$ at
$z=d$. These surfaces are also assumed to be impermeable and
stress-free. The magnetic field satisfies $B_x=B_y=0$ at $z=0$ and
$z=d$, which corresponds to a vertical magnetic field boundary
condition. 

\par The governing equations are those of non-ideal, compressible
magnetohydrodynamics, with an appropriate equation of state
for a perfect gas \citep[see, e.g.][]{BHPW08}. We non-dimensionalise
this system \citep[see][for more details]{BHPW08}, scaling
all lengths by $d$, whilst all time-scales are expressed in terms
of an (isothermal) acoustic travel time, $d/(R_*T_0)^{1/2}$. Hence,
the fluid velocity, $\mathbf{u}$, is scaled in terms of the
unperturbed isothermal sound speed at the upper surface,
$(R_*T_0)^{1/2}$. The temperature, $T$, and density, $\rho$, are scaled by
$T_0$ and $\rho_0$ respectively (where $\rho_0$ is the density at the
upper surface in the absence of convection). We scale the magnetic
field, $\mathbf{B}$, by $\left(\mu_0\rho_0R_*T_0\right)^{1/2}$. This
choice of scaling for the magnetic field implies that the Alfv\'en
speed at the upper surface (like the fluid velocity) is also expressed
in terms of the isothermal sound speed. With these scalings, the
governing equations become 

\begin{eqnarray}
&&\frac{\partial \rho}{\partial t}=- \nabla \cdot \left(\rho
\mathbf{u}\right),\\ \nonumber \\ 
&&\frac{\partial}{\partial t}\left(\rho \mathbf{u}\right)=- \nabla
\left(P + |\mathbf{B}|^2/2\right) +\theta(m+1)\rho\mathbf{\hat{z}}\\
\nonumber&& \hspace{0.65in} + \nabla \cdot \left( \mathbf{BB} - \rho \mathbf{uu} +
\kappa \sigma \mathbf{\tau}\right), \hspace{0.4in} \\ \nonumber \\
&&\frac{\partial T}{\partial t}= -\mathbf{u}\cdot\nabla T -
\left(\gamma -1\right)T\nabla \cdot \mathbf{u} +
\frac{\kappa\gamma}{\rho}\nabla^2 T \\ \nonumber &&\hspace{0.35in}+
\frac{\kappa(\gamma-1)}{\rho}\left(\sigma \tau^2/2 + \zeta_0|\nabla
\times \mathbf{B}|^2\right),\\ \nonumber \\
&&\frac{\partial \mathbf{B}}{\partial t}=\nabla \times \left( \mathbf{u}
\times \mathbf{B} -  \kappa \zeta_0 \nabla \times \mathbf{B} \right),
~~\nabla \cdot \mathbf{B} = 0.
\end{eqnarray}

\noindent The pressure, $P$, satisfies $P=\rho T$, whilst the
components of the stress tensor, $\mathbf{\tau}$ are
given by

\begin{equation}
\tau_{ij}= \frac{\partial u_i}{\partial x_j}+\frac{\partial
 u_j}{\partial x_i} - \frac{2}{3}\frac{\partial u_k}{\partial
 x_k}\delta_{ij}.
\end{equation}

\noindent Note that these equations have a simple equilibrium
solution, corresponding to a hydrostatic polytropic layer, in which
$\mathbf{u}=\mathbf{B}=\mathbf{0}$, $T=1+\theta z$ and
$\rho=T^m$. 

\begin{table}
\caption{The non-dimensional parameters}
\label{table1}
\centering
\begin{tabular}{cccc}
\hline\hline 
Parameter & Definition & Values used \\ \hline
$\gamma$ & $c_P/c_V$ & $5/3$\\
$m$ & $g d/R_*\Delta T - 1$ & $1.0$\\
$\theta$ & $\Delta T/T_0$ & $4.0$\\
$\kappa$ & $K/\rho_0 d c_P(R_*T_0)^{1/2}$ & $0.0438$ \\
$\zeta_0$ & $\eta \rho_0 c_P /K$ & $0.2$, $0.1$ or $0.05$\\
$\sigma$ & $\mu c_P/K$ & $0.1$ \\
\hline
\end{tabular}
\end{table}

\subsection{Model parameters}
Various non-dimensional parameters appear in this model. These are
defined in Table~\ref{table1}. Here, $\gamma=5/3$ is the ratio of the
specific heat capacities, $m=1.0$ is the polytropic index, whilst $\theta=4.0$
is a measure of the thermal stratification. This choice of parameters implies
that the layer is superadiabatically-stratified, with the temperature
varying by a factor of $5$ across the layer. The parameter $\kappa$ is
a non-dimensionalised thermal diffusivity, $\sigma$ is the Prandtl
number, whilst $\zeta_0$ represents the ratio of the magnetic to the
thermal diffusivity at the top of the layer. To interpret these
coefficients, it is convenient to introduce the Reynolds number

\begin{equation}
\mathcal{R}e = \frac{\rho_{mid}U_{rms}}{\kappa \sigma}
\end{equation}

\noindent (where $U_{rms}$ is the rms velocity and $\rho_{mid}$ is the
horizontally-averaged mid-layer density) and the magnetic
Reynolds number

\begin{equation}
\mathcal{R}m = \frac{U_{rms}}{\kappa \zeta_0}.
\end{equation}

\noindent Note that the depth of the layer, which equals unity in these dimensionless units, has been used as the characteristic lengthscale in the definitions of these Reynolds numbers. With the parameter values that are given in
Table~\ref{table1}, $\mathcal{R}e\approx 420$ for the fully-developed hydrodynamic flow, which implies that the
convection is highly turbulent. We have chosen three values for $\zeta_0$, 
which imply that $\mathcal{R}m\approx 70$ (for $\zeta_0=0.2$), $\mathcal{R}m\approx 140$ (for $\zeta_0=0.1$)  or
$\mathcal{R}m\approx 280$ (for $\zeta_0=0.05$). Note that the magnetic
Prandtl number, $Pm = \mathcal{R}m/\mathcal{R}e$ is either $0.17$, $0.33$ or
$0.67$. As described in the Introduction, $Pm \ll 1$ throughout the
solar convection zone. Although that parameter regime cannot be reached
in direct numerical simulations, we have at least ensured that $Pm < 1$
in all cases. The exploration of the ``low'' $Pm$ regime is one of the
novel aspects of this study. 

\begin{figure}
\centering
\resizebox{\hsize}{!}{\includegraphics{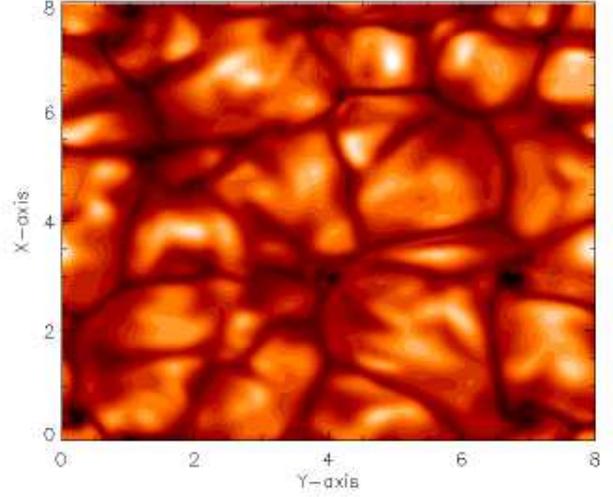}}
\caption{The temperature distribution in a horizontal plane just
  below the upper surface of the domain, at $t=0$. Brighter contours
  correspond to regions of warmer fluid.}  
\label{fig1}
\end{figure}

\subsection{Initial conditions}
All of the calculations that are considered in this paper are based
upon a particular hydrodynamic flow (as determined by the parameters
that are given in Table~\ref{table1}). This is evolved in time until
the convection has reached a statistically-steady state. A snapshot of
the resulting flow is shown in Figure~\ref{fig1}. This plot shows the
temperature distribution in a horizontal plane just below the upper
surface of the computational domain. Bright contours correspond to
warm upflows, whilst dark contours correspond to cooler convective
downflows. The flow is characterised by a time-dependent
pattern of granular convection, reminiscent of the one observed at the
solar surface. Once the convection has reached a statistically-steady state, we
insert a magnetic field into the flow. In what follows, we define this
insertion time to be $t=0$. This magnetic field takes the form of a 
weakly twisted, horizontal magnetic flux tube that is aligned with the
$y$-axis. The radius of the tube, $R$, is assumed to be $0.1$ (which
is $10\%$ of the depth of the domain) in most cases, although a small
number of calculations are carried out with a wider tube ($R=0.15$)
. The axis of the flux tube is centred at  $x=4.0$ and $z=0.75$. The
three components of this field are given by 

\begin{eqnarray}
B_y &=& B_0 \mathrm{e}^{\left[-\left(r/R\right)^2\right]}\\ \nonumber
B_x &=& 0.4 (z-0.75)B_y\\ 
B_z &=& -0.4 (x-4.0)B_y,\nonumber 
\end{eqnarray}
where $r=\sqrt{(x-4.0)^2 + (z-0.75)^2}$ and $B_0$ is a
free parameter that can be adjusted to control the strength of the
magnetic field. The factor of $0.4$ in the definitions of $B_x$ and $B_z$ implies that the initial flux tube is weakly twisted. The effect of increasing the twist of the tube (by increasing this factor) is briefly discussed in the next section.

\par The strength of the magnetic field is a free parameter at this
stage. Recalling that we are working with dimensionless variables, it
is important to relate the value of $B_0$ to the properties of the
flow. We define the plasma $\beta$ to be the ratio of the gas
pressure in the unperturbed polytrope at the initial depth of the tube
(i.e. at $z=0.75$) to the magnetic pressure. The three cases that are
considered in detail in this paper are $B_0=0.96$,
$B_0=2.53$ and $B_0=4.62$, which correspond to $\beta=35$, $\beta=5$
and $\beta=1.5$ respectively. It is also of interest to relate the
initial magnetic energy density of the field to the kinetic energy
density of the convection. At a depth of $z=0.75$, a value of $B_0=1.25$ would
give a peak field strength that would be in approximate equipartition
with the kinetic energy of the local convective motions. However, given that the motion is
dominated by strong convective downflows, it makes more sense to define an
equipartition field strength based upon the mean kinetic energy of the
convective downflows at this depth. This implies an equipartition
value of $B_{eq}=0.96$, which (in turn) implies that the three cases that are
considered in detail in this paper correspond to $B_0\approx B_{eq}$,
$B_0 \approx 2.6 B_{eq}$ and $B_0 \approx 4.8 B_{eq}$. In this sense, this range of
values for $B_0$ is similar to those considered in previous
studies. 

\par Magnetic fields of this strength will exert a dynamically-significant Lorentz force
upon the flow. For a weakly twisted flux tube of this form, the $x$
and $z$ components of the magnetic field are much smaller than the
$B_y$ component. As a result of this, the dominant component of the Lorentz force is
due to radial gradients in the magnetic pressure
which tend to force fluid away from the tube axis. To compensate for
this magnetic pressure, we reduce the gas pressure inside the magnetic
flux tube at $t=0$ in such a way that the total (gas plus magnetic) pressure
distribution inside the tube matches the gas pressure distribution
just before the field was introduced. Although the tube is not in
perfect force balance with its surroundings, this simple method produces a
very good approximation to such a state. Following \citet{CSM07} we
also modify the local velocity field within the flux tube so as to
minimise convective perturbations to the flux tube during the early
stages of evolution. Using the notation that we adopted for the magnetic
field, we set 

\begin{equation}
\mathbf{u}_{new}=\left(1-\mathrm{e}^{\left[-\left(r/R\right)^2\right]}\right)\mathbf{u}.
\end{equation}
\noindent As described in \citet{CSM07}, this corresponds to a
velocity field that vanishes along the axis of the tube, smoothly
matching up to the original velocity field near the edge of the tube. 

\par Having specified the gas pressure inside the flux tube, we are free to
specify one other thermodynamic variable. When considering buoyancy
instabilities, it is natural to discuss the entropy of the flux
tube. We specify an initial entropy distribution $S_{new}$ that is maximal (and
uniform) along the tube's axis, but matches smoothly onto the
background entropy profile, $S$, at its outer radius. Hence we set

\begin{equation}
S_{new} = S\left(1-\mathrm{e}^{\left[-\left(r/R\right)^2\right]}\right)+S_0\mathrm{e}^{\left[-\left(r/R\right)^2\right]}. 
\end{equation} 

\noindent Normalising the entropy by $S_{mean}(0.75)$, the horizontally-averaged
entropy at $z=0.75$, we consider two different cases. In the lower entropy case, the peak entropy along the axis of the tube is given by $S_0=1.26S_{mean}(0.75)$. This value of $S_0$ is comparable to the peak entropy in the convective upflows at this depth. In the higher entropy case, we choose a value for $S_0$ that is $50\%$ higher than this, i.e. $S_0=1.9S_{mean}(0.75)$. 
Note that we have also considered some other cases (not shown in this 
paper) in which the entropy was uniform in the region $r<0.1$. In fact, these
simulations evolved in a very similar way to the radially varying
cases. This is presumably due to the fact that thermal diffusion
rapidly smooths out steep gradients in temperature (which, in turn,
smooths out any sharp radial gradients in the entropy
distribution). Given that the evolution of these simulations seems to be rather
insensitive to the precise radial distribution of entropy in the
vicinity of the flux tube, only the radially non-uniform cases will be
discussed in this paper. 
\begin{table}
\caption{A summary of the simulations}
\label{table2}
\centering
\begin{tabular}{ccccccc}
\hline\hline 
Simulation & $S_0$/$S_{mean}$ & $R$ & $\beta$ & $\mathcal{R}e$
& $\mathcal{R}m$ & $Pm$ \\ \hline
L1 & $1.26$ & $0.1$ & $1.5$ & $420$ & $140$ & $0.33$ \\
L2 & $1.26$ & $0.1$ & $5.0$ & $420$ & $140$ & $0.33$ \\
L3 & $1.26$ & $0.1$ & $35.0$ & $420$ & $140$ & $0.33$ \\
L4 & $1.26$ & $0.1$ & $1.5$ & $420$ & $70$ & $0.17$ \\
L5 & $1.26$ & $0.1$ & $5.0$ & $420$ & $70$ & $0.17$ \\
L6 & $1.26$ & $0.1$ & $35.0$ & $420$ & $70$ & $0.17$ \\
L7 & $1.26$ & $0.1$ & $1.5$ & $420$ & $280$ & $0.67$ \\
L8 & $1.26$ & $0.1$ & $5.0$ & $420$ & $280$ & $0.67$ \\
L9 & $1.26$ & $0.1$ & $35.0$ & $420$ & $280$ & $0.67$ \\
WT1 & $1.26$ & $0.15$ & $1.5$ & $420$ & $140$ & $0.33$ \\
WT2 & $1.26$ & $0.15$ & $1.5$ & $420$ & $70$ & $0.17$ \\
WT3 & $1.26$ & $0.15$ & $1.5$ & $420$ & $280$ & $0.67$ \\
H1 & $1.9$ & $0.1$ & $1.5$ & $420$ & $140$ & $0.33$ \\
H2 & $1.9$ & $0.1$ & $5.0$ & $420$ & $140$ & $0.33$ \\
H3 & $1.9$ & $0.1$ & $35.0$ & $420$ & $140$ & $0.33$ \\
E1 & - & $0.1$ & $1.5$ & $420$ & $140$ & $0.33$ \\
E2 & - & $0.15$ & $1.5$ & $420$ & $140$ & $0.33$\\
\hline
\end{tabular}
\end{table}
\par Table~\ref{table2} summarises the numerical simulations in this study. Most of the simulations (L1--L9) correspond to the lower entropy, thin tube ($R=0.1$) cases, for different values of $\beta$ and different values of $\mathcal{R}m$.  H1, H2 and H3 illustrate the effects of increasing the initial peak entropy within the thin flux tube (but are otherwise identical to cases L1, L2 and L3). WT1, WT2 and WT3 represent three calculations that were carried out with a wider flux tube, using the lower entropy initial conditions. Finally, in the E1 and E2 simulations, the velocity field, the gas pressure and the entropy distribution are exactly as they were before the field was
introduced. Thus, no modification is made to the convection at $t=0$. This implies that the Lorentz force will significantly perturb the flow during the early stages of evolution. The rationale for these particular calculations will be discussed later in the paper.   

\par
\begin{figure}
\resizebox{\hsize}{!}{\includegraphics{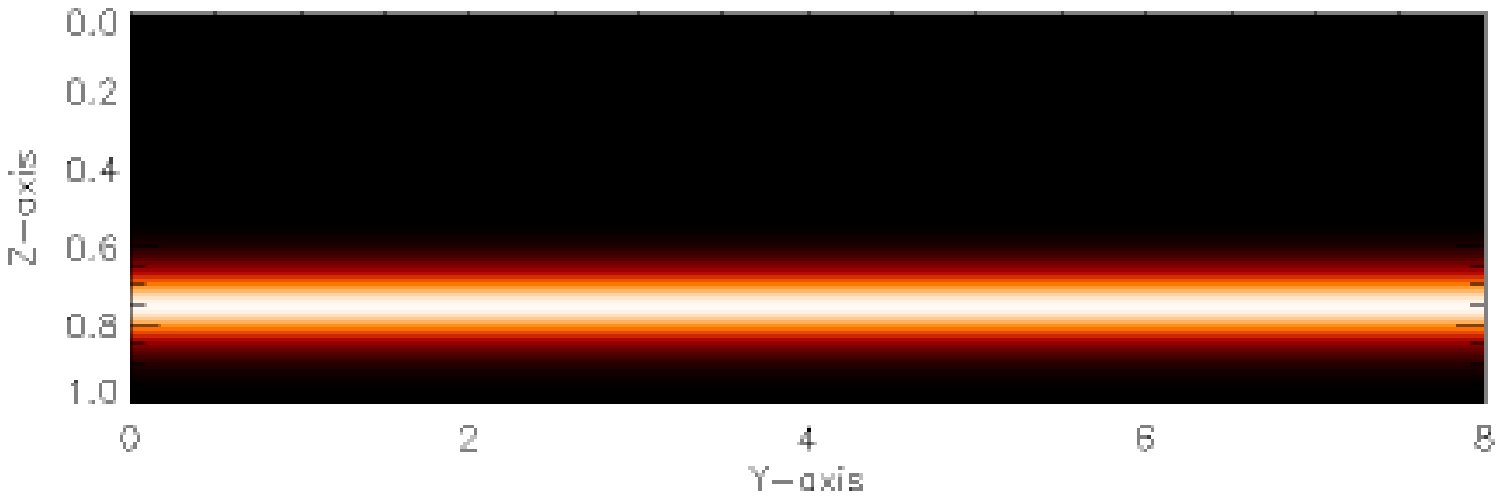}}
\resizebox{\hsize}{!}{\includegraphics{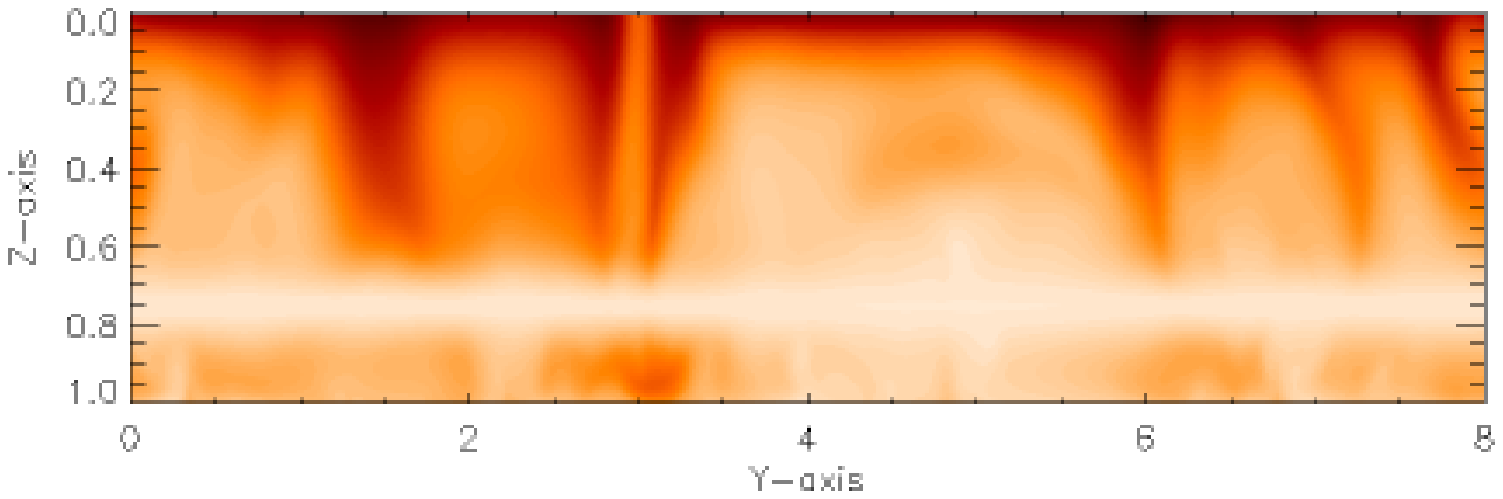}}
\resizebox{\hsize}{!}{\includegraphics{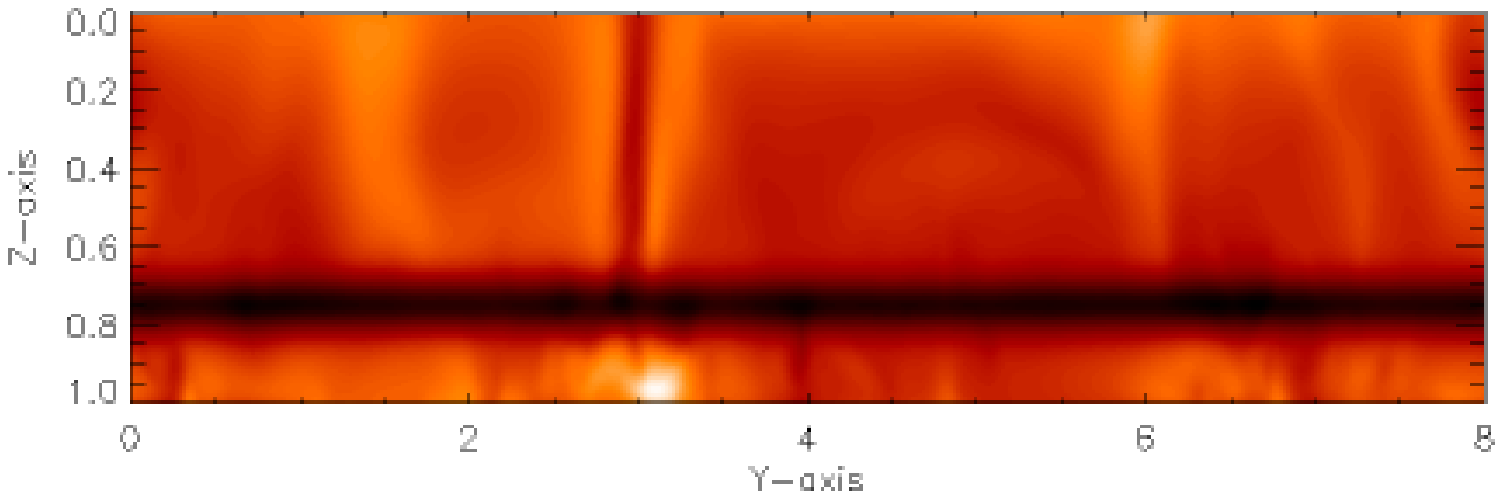}}
\caption{Vertical slices through the $x=4.0$ plane, showing contours
  of $B_y$ (top), entropy (middle) and the density perturbation (bottom)
  for the L1 case at $t=0$.} 
\label{fig2}
\end{figure}

 We illustrate the initial condition for one of the low entropy
cases (L1) in Figure~\ref{fig2}, which shows the distribution
of various quantities in the $x=4.0$ plane. The upper plot shows
contours of constant $B_y$. The localised nature of this initial flux
tube is clearly apparent. The middle plot in Figure~\ref{fig2} shows
contours of constant entropy. It is clear from this plot that the peak
entropy in the flux tube is comparable to the peak entropy in the convective
upflows. We would expect a tube with this entropy distribution to be
buoyant relative to its surroundings. Finally, the lower part of
Figure~\ref{fig2} shows the density perturbation in this plane. To
obtain this plot, we have subtracted off the polytropic background
stratification in order to highlight the variations in $\rho$ due to the magnetic flux tube. As expected, the density perturbation in this strong field ($\beta=1.5$) case takes its
minimum value within the flux tube. Here, the minimum value of $\rho$
is approximately $40\%$ of the mean density at this depth. The
associated reduction in density is smaller in the $\beta=5$ case, and
almost non-existent in the $\beta=35$ case. This simply reflects the
fact that the magnetic field is weaker in the higher $\beta$ cases.

\par To evolve these simulations, we use a
well tested mixed pseudospectral/finite difference code. All
horizontal derivatives are evaluated in Fourier space, using standard
Fast Fourier Transform (FFT) libraries. Fourth-order finite
differences are used to calculate vertical derivatives. The
time-stepping is carried out using a third-order Adams-Bashforth
scheme, with a variable time-step. The code is parallelised using
MPI. All simulations are carried out on a $512 \times 512 \times 96$ mesh.

\section{Results}

\subsection{Varying the initial conditions}

\begin{figure}
\resizebox{\hsize}{!}{\includegraphics{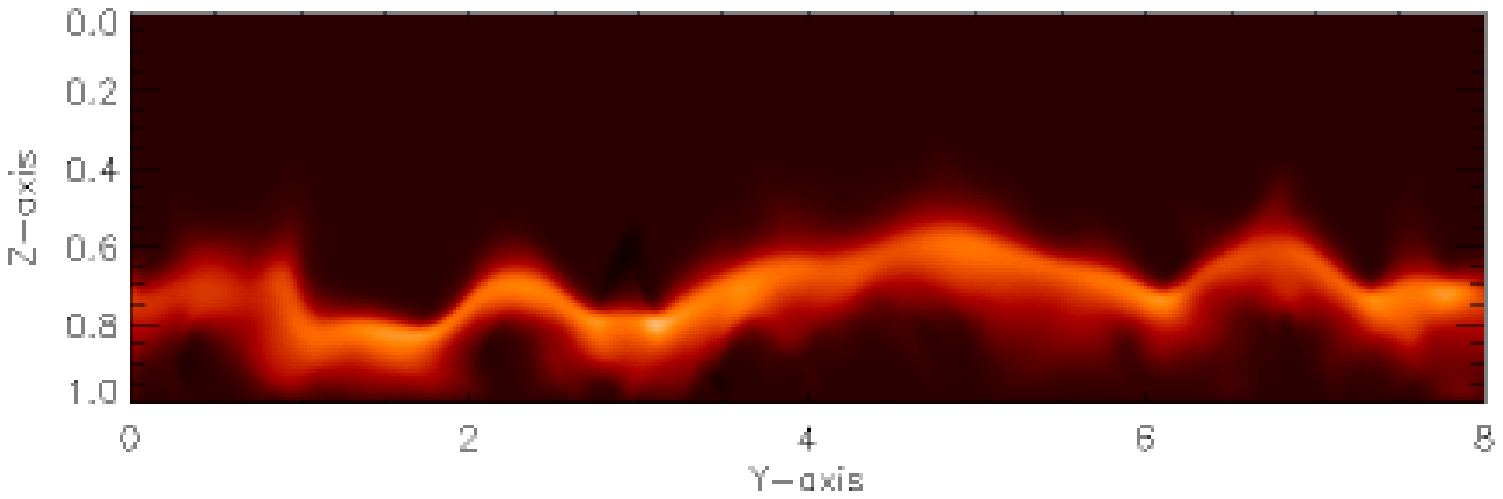}}
\resizebox{\hsize}{!}{\includegraphics{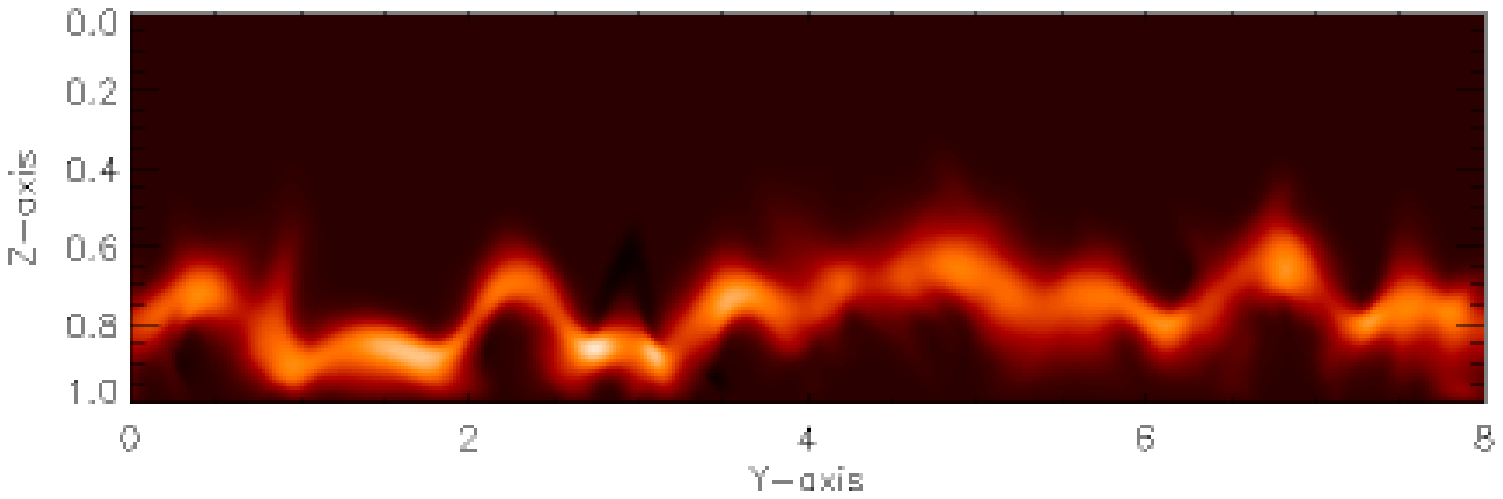}}
\resizebox{\hsize}{!}{\includegraphics{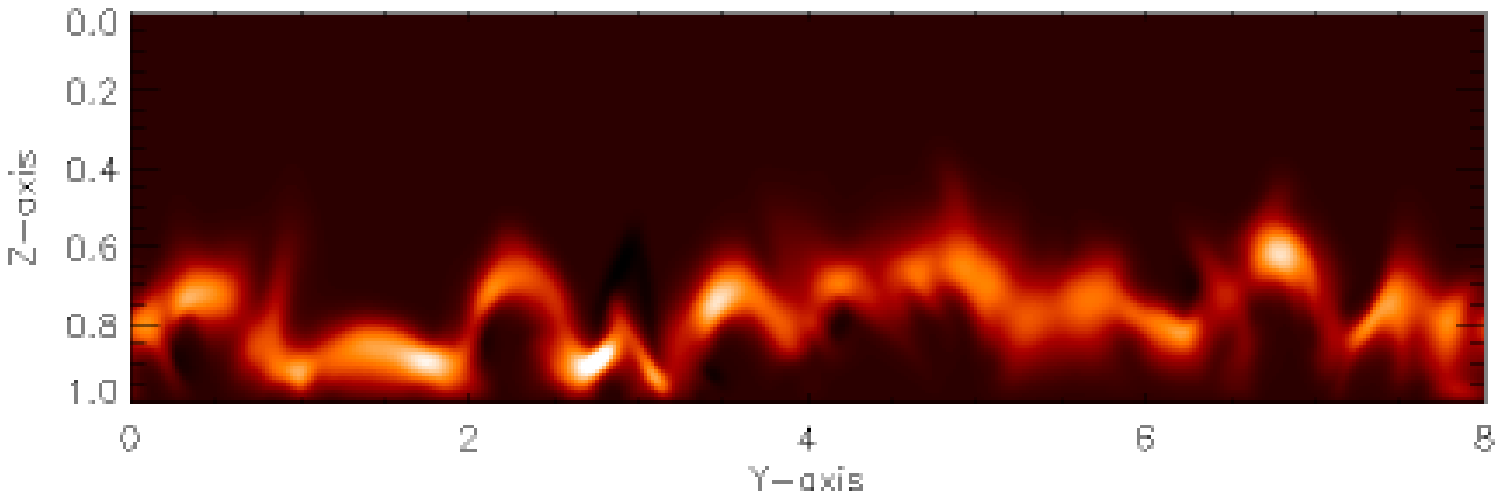}}
\caption{Vertical slices through the $x=4.0$ plane, showing contours
  of $B_y/B_0$ at $t=0.36$ for three different cases, L1 ($\beta=1.5$, top), L2 ($\beta=5.0$, middle) and L3 ($\beta=35$, bottom). In each case, the contours are evenly spaced in the range $-0.13 \le B_y/B_0 \le 0.9$.}
\label{fig3}
\end{figure}

We first consider cases L1, L2 and L3, which correspond to the lower
entropy simulations at $\mathcal{R}m\approx 140$, for $\beta=1.5$, $\beta=5$
and $\beta=35$ respectively. In all of these cases, convective motions
play a significant role in determining the early evolution of the system, disrupting
the flux tube by pulling out loops of magnetic field that are
then advected around the domain. The early stages of convective
disruption are illustrated in Figure~\ref{fig3}, which shows contours
of $B_y/B_0$ in the $x=4.0$ plane for each of these cases at
$t=0.36$. Normalising the magnetic field by $B_0$ allows us to compare
cases with different values of $\beta$. Defining the convective
turnover time, $\tau_{conv}$, to be the inverse of the rms velocity,
$\tau_{conv} \approx 1.56$ in these dimensionless units. Hence
$t=0.36$ corresponds to a small fraction of the turnover
time. However, even after this short time period, we see that the flux
tube has been perturbed in a significant way by the convective
flows. This convective disruption is most apparent in the weaker field
(higher $\beta$) cases, where small-scale perturbations to the
magnetic field have produced a magnetic field with an undulating {\it
  sea-serpent} configuration \citep[as observed in previous
calculations, e.g.][]{CSM07}. Only in the $\beta=1.5$ case does the
field appear to be strong enough to partially resist the convective disruption on this time-scale. Obviously stronger fields possess a higher level of magnetic tension, which makes it more
difficult for the surrounding convective motions to perturb the magnetic
flux tube.

\par
\begin{figure*}
\centering
\includegraphics[width=9cm]{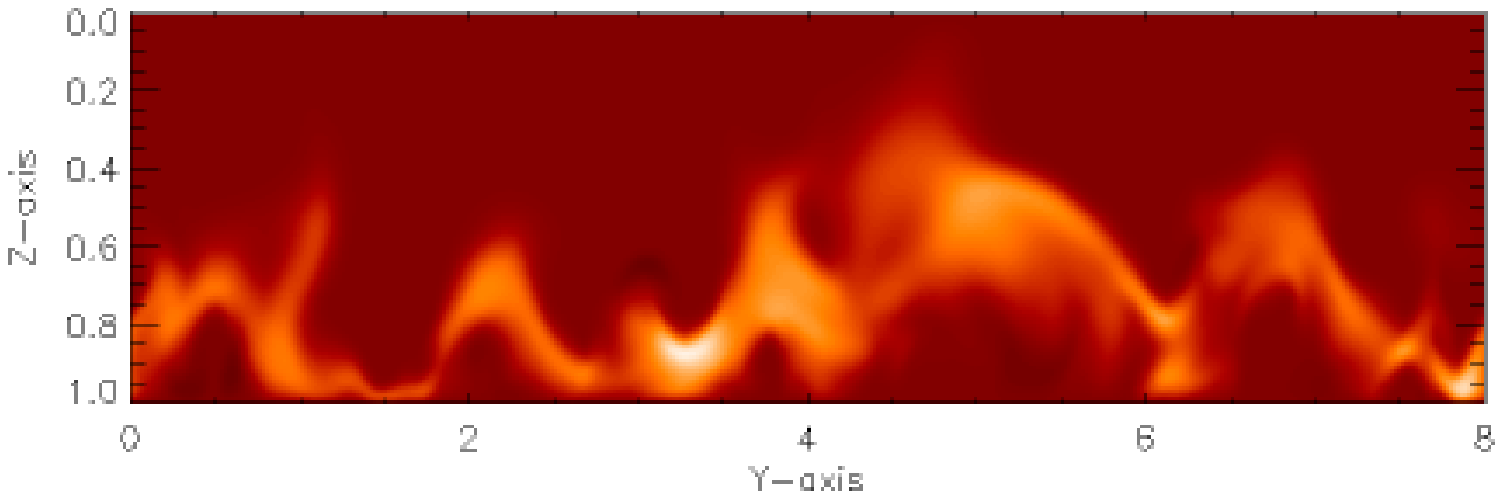}
\includegraphics[width=9cm]{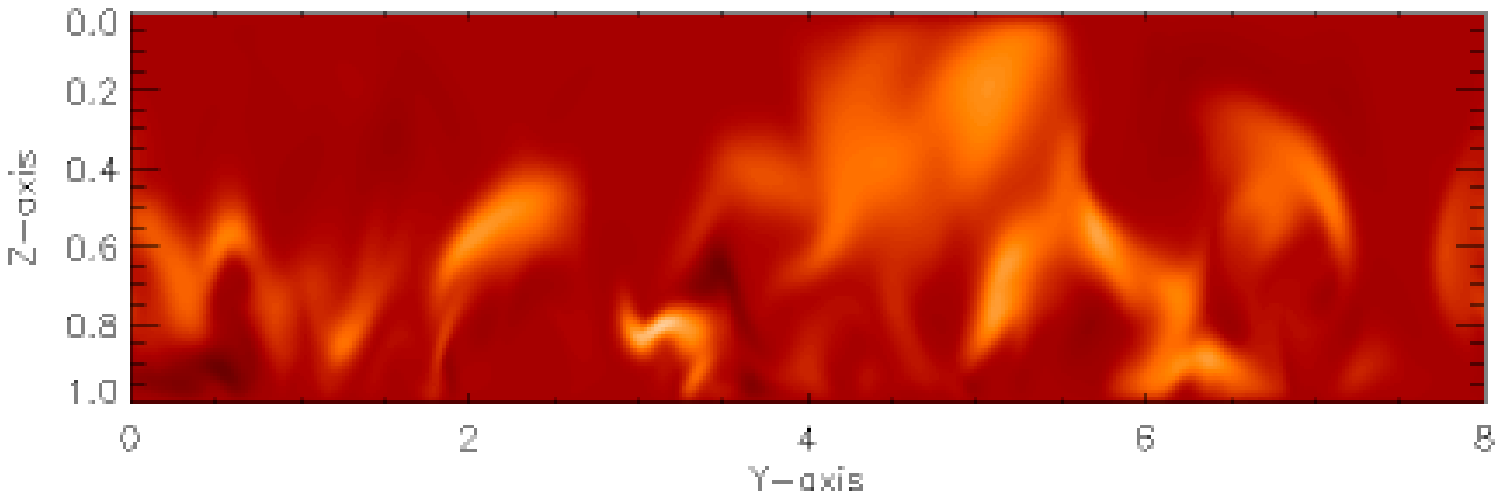}
\includegraphics[width=9cm]{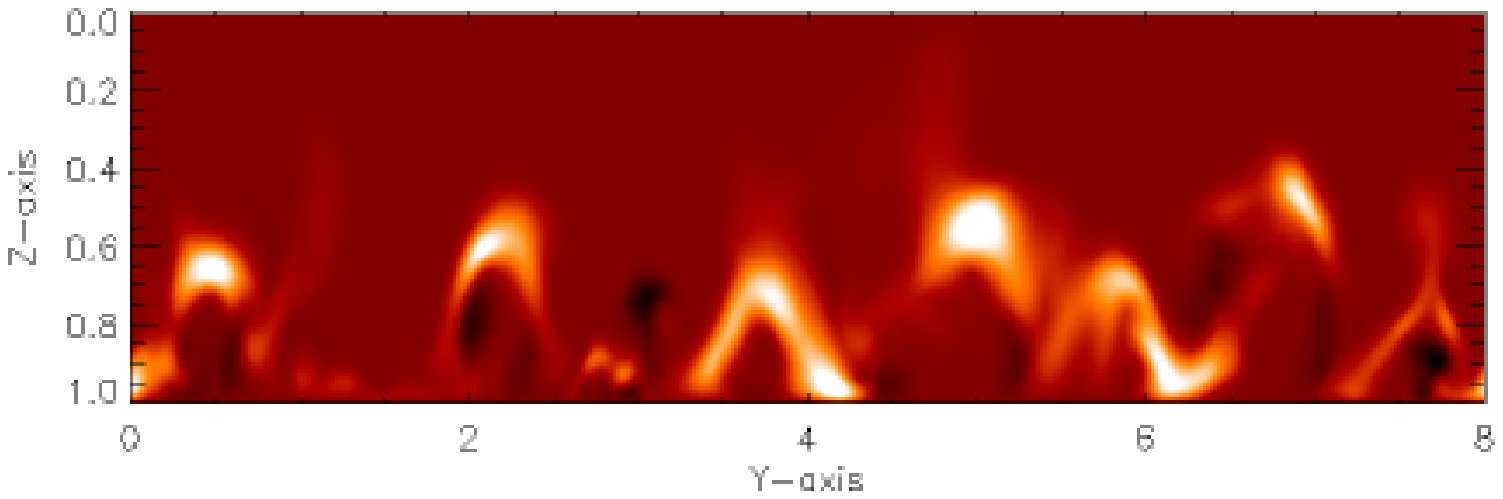}
\includegraphics[width=9cm]{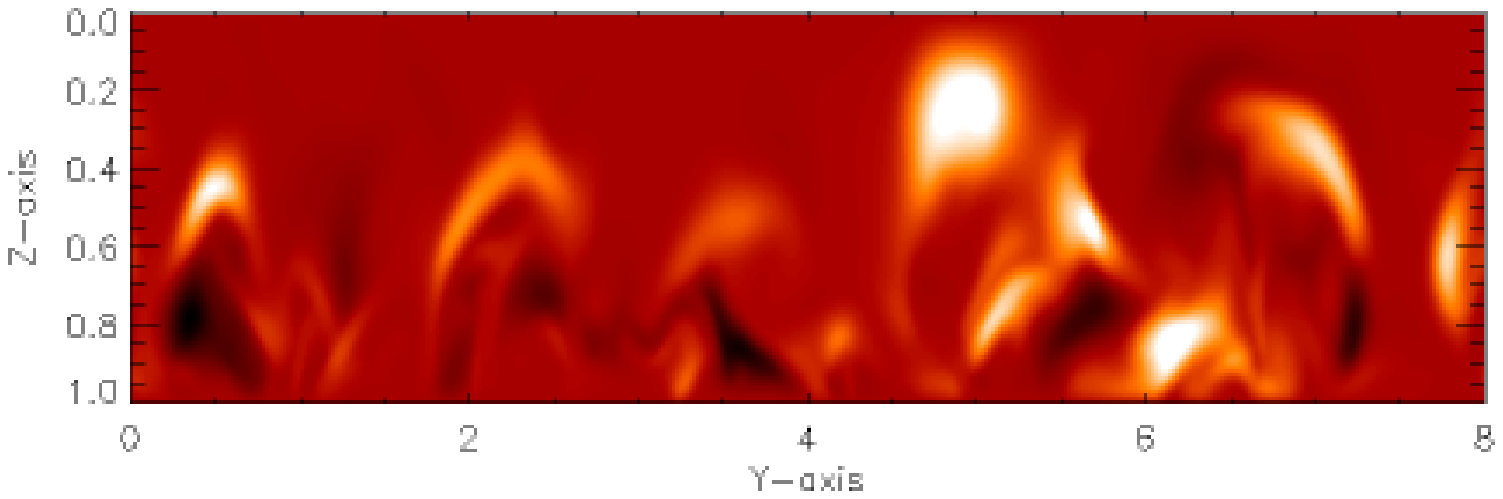}
\caption{Contours of $B_y/B_0$ in the $x=4.0$ plane. Top: L1 ($\beta=1.5$) at $t=0.80$ (left) and $t=1.59$ (right). Bottom: The same plots for the L3 ($\beta=35$) case. In the $t=0.80$ plots, the contours are evenly spaced in the range $-0.25 \le B_y/B_0 \le 0.45$, whilst the contour range is $-0.22 \le B_y/B_0 \le 0.26$ in the $t=1.59$ plots.}
\label{fig4}
\end{figure*}

 The subsequent evolution of cases L1 and L3 is illustrated in Figure~\ref{fig4}, which shows the contours
of $B_y/B_0$ in the $x=4.0$ plane, for each case, at $t=0.80$ and
$t=1.59$. At these later times, the differences between the stronger
field and the weaker field case have become more pronounced. Focusing
initially upon the weaker field case (L3), it is clear that the flux
tube has been further disrupted by the convective motions. In the
$t=0.80$ plot, the upflows and downflows in the lower part of the
domain have produced many loop-like structures, with several
well-defined regions of negative $B_y$ (which correspond to the dark
contours in Figure~\ref{fig4}). Given the relatively weak magnetic
fields that are present at this stage of the simulation (even the peak
horizontal field strength is now well below the equipartition value
that was discussed in the previous section), we would expect the
magnetic field to play a relatively passive role in the dynamics. The
efficiency with which the convective motions have advected the
magnetic flux around the domain would appear to be consistent with
this idea. We can see from the $t=1.59$ plot for the L3 case that a
strong upflow in the vicinity of $y=5$ has advected some horizontal
magnetic flux into the upper layers of the computational
domain. Although convective disruption does also play a role in the
stronger field case (L1), there are fewer loop-like structures in the
magnetic field distribution at $t=0.80$, which suggests that the
magnetic flux tube maintains a greater degree of coherence than in the
weaker field case. Furthermore, as illustrated in the $t=1.59$ plot, a
much broader concentration of magnetic flux rises towards the surface
regions in this case than it does in the corresponding weaker field
calculation. Even though the peak field strength in this emerging flux
concentration is a relatively small fraction of its initial value,
$B_0$, it is still comparable to the equipartition
value, which suggests that it should still exert a
dynamically-significant Lorentz force upon the flow. We shall discuss the flux emergence pattern for the L1 case in more detail in the next subsection. 

\par Figures~\ref{fig3} and~\ref{fig4} clearly show that there is some dependence upon $\beta$ in these simulations. The evolution of the flux tube in the weakest field case (L3) appears to be largely (possibly completely) determined by the surrounding convective motions. However, further quantitative analysis is needed in order to determine whether or not magnetic buoyancy is influencing the evolution of the flux tube in the lower $\beta$ cases. In order to investigate this possibility, we have traced the evolution of the unsigned emerging magnetic flux at $z=0$. This is given by

\begin{equation}
\Phi(t) = \int_0^8\int_0^8 |B_z(z=0,t)|\,\mathrm{d}x\,\mathrm{d}y.
\end{equation}

\begin{figure}
\resizebox{\hsize}{!}{\includegraphics{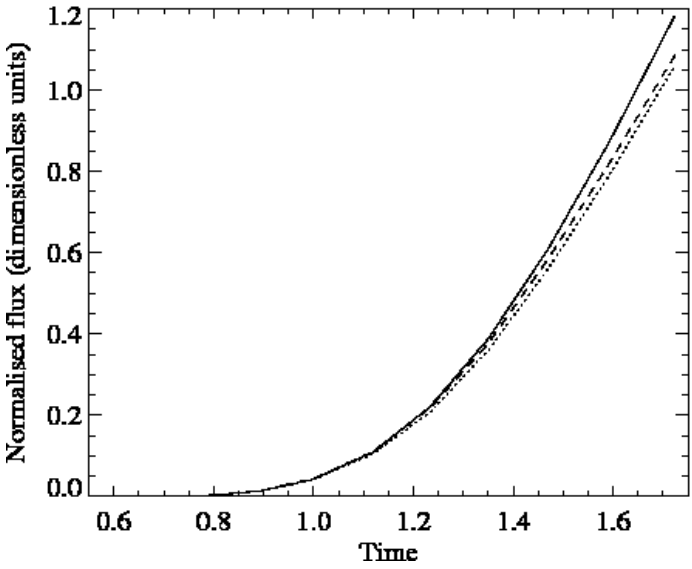}}
\caption{The magnitude of the total unsigned magnetic flux, $\Phi(t)$,
  at $z=0$, plotted against time for simulations L1, L2 and L3 (the
  lower entropy, thin tube, $\mathcal{R}m\approx 140$ cases). The solid line
  corresponds to L1 ($\beta=1.5$), the dotted line corresponds to
  L2 ($\beta=5$), whilst the dashed line corresponds to L3 ($\beta=35$). In each case, the flux has been normalised by the initial magnetic flux in the $y$-direction.}
\label{fig5}
\end{figure}

\noindent To compare cases at different values of $\beta$, we
normalise this quantity by the initial flux in the $y$-direction. The
results of this quantitative analysis for L1, L2 and L3 are shown in
Figure~\ref{fig5}. Cases L2 and L3 both give very similar flux
emergence rates, which indicates that the flux emergence in each case
is driven almost entirely by the convective upflows. The marginal
difference between the flux emergence rates in these two cases would
then simply reflect the fact that stronger fields are more resistant
to advection by the convective motions. However, the slightly higher flux emergence rates in simulation L1 can only be explained if magnetic buoyancy is enhancing the flux emergence rate in this case. We therefore conclude that magnetic buoyancy is playing a role in the evolution of the flux tube in this lower $\beta$ (stronger field) calculation, helping the convective upflows to advect dynamically-significant concentrations of magnetic flux into the upper layers of the domain.  

\par
\begin{figure}
\resizebox{\hsize}{!}{\includegraphics{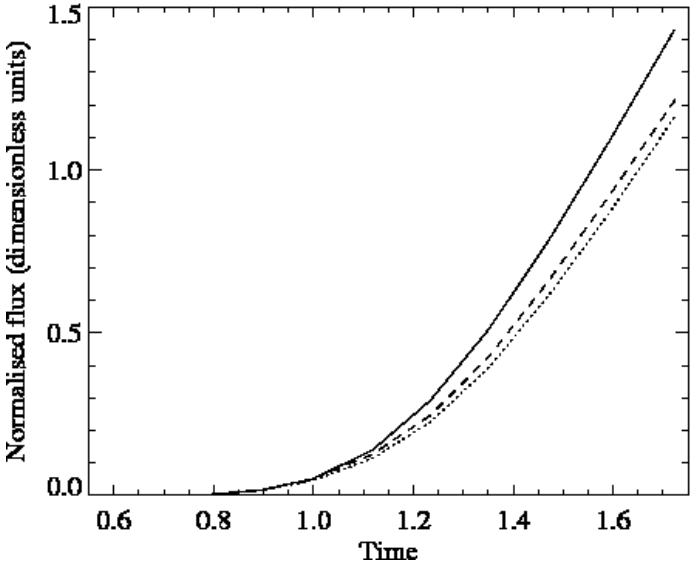}}
\caption{As figure~\ref{fig5}, but this time for the high entropy
  cases at $\mathcal{R}m\approx 140$ (H1, H2 and H3). The solid line
  corresponds to H1 ($\beta=1.5$), the dotted line corresponds to
  H2 ($\beta=5$), whilst the dashed line corresponds to H3 ($\beta=35$).}
\label{fig6}
\end{figure}

 So far only the strength of the imposed field has been
varied. As indicated in Table~\ref{table2}, cases H1, H2 and H3
adopt the same values of $\beta$ and $\mathcal{R}m$ as L1, L2 and L3 (respectively), but this time with a 
higher peak entropy within the initial flux tube. Figure~\ref{fig6} shows the flux emergence as a function of time for these cases. Comparing this plot with Figure~\ref{fig5}, it is clear that there is generally an enhanced rate of flux emergence in these higher entropy calculations. This is to be expected given that the higher entropy flux tube should be more buoyant than those considered in cases L1, L2 and L3. However, it is again notable that there is only a modest dependence upon $\beta$ in the weaker field cases, H2 and H3, which exhibit very similar flux emergence rates. As in the lower entropy case, there is again a slightly higher rate of flux emergence in the H1 ($\beta=1.5$) case. As before, this suggests that magnetic buoyancy plays a role in the stronger field calculation. We can conclude from this that the enhancement of the flux emergence rate in the stronger field case is a robust result that does not depend critically upon the choice of the initial entropy distribution along the axis of the flux tube. Some simulations with even higher values of the initial entropy were considered. However the emerging flux in these cases tended to be associated with regions of exceptionally high temperature (something that is not observed in the Sun), so these calculations were not pursued in any detail.  

\subsection{Varying the flux tube geometry}

\begin{figure}
\resizebox{\hsize}{!}{\includegraphics{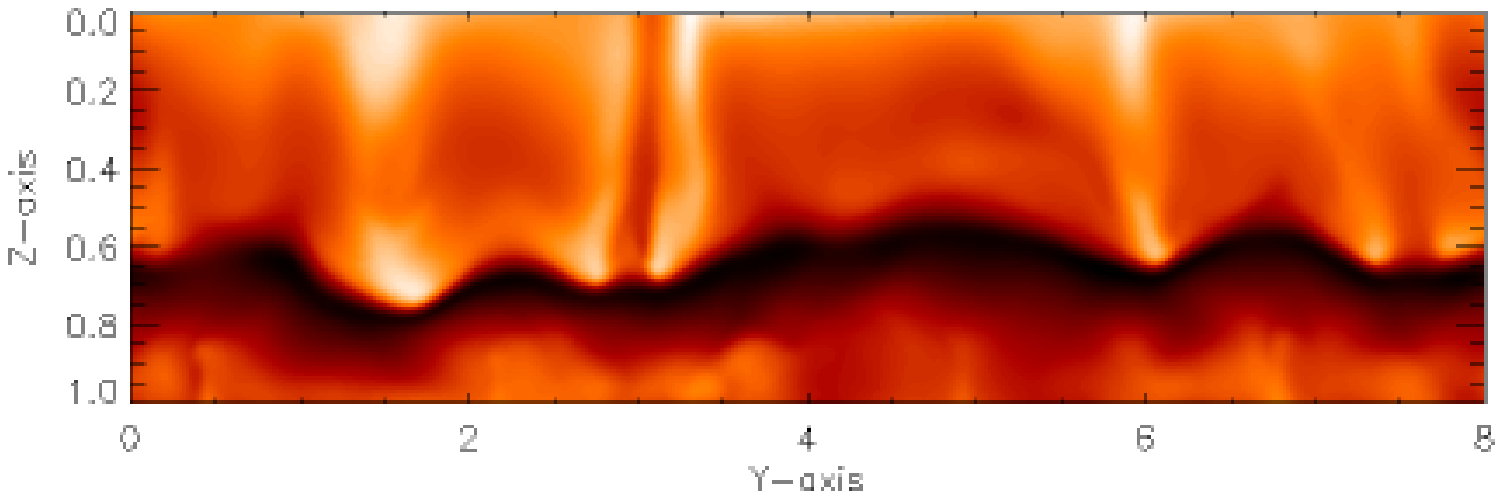}}
\resizebox{\hsize}{!}{\includegraphics{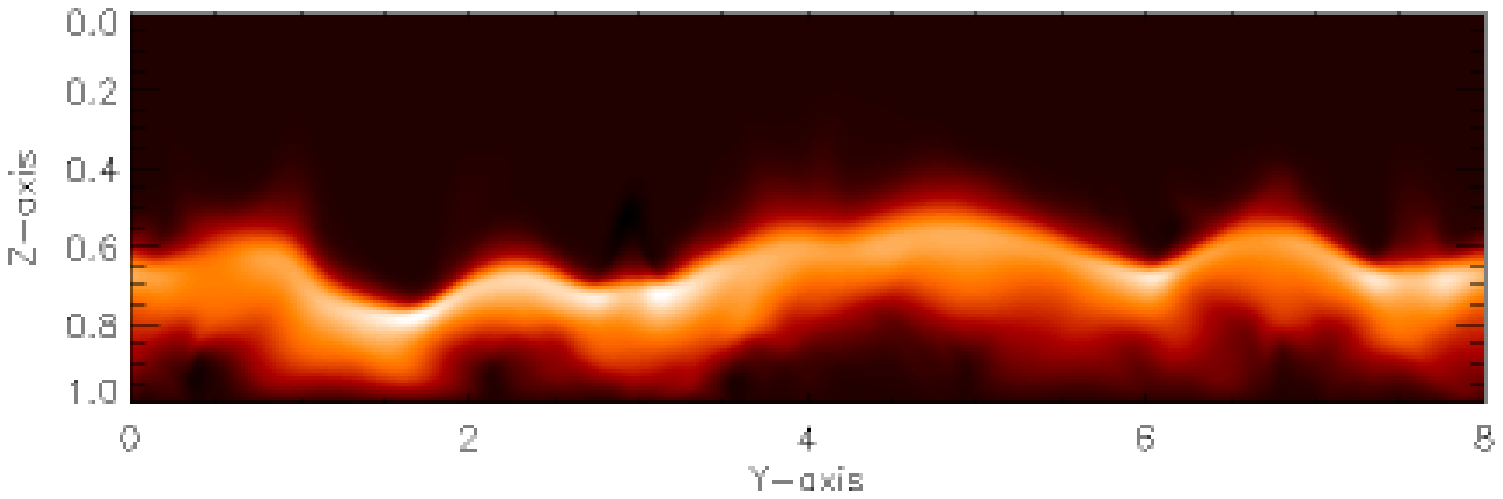}}
\resizebox{\hsize}{!}{\includegraphics{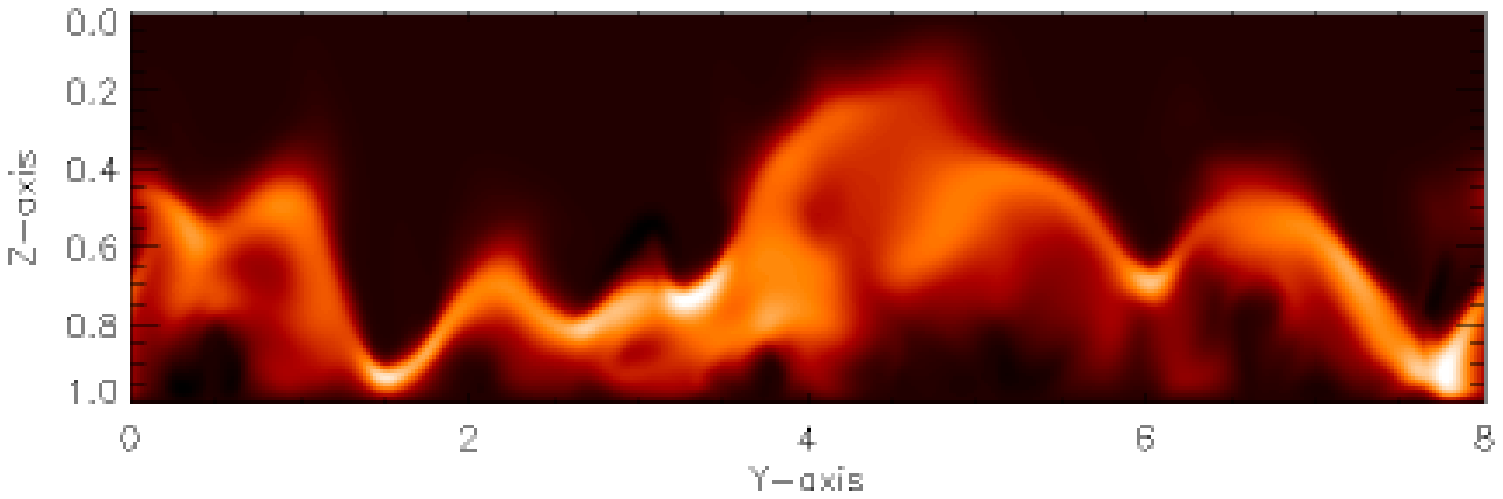}}
\caption{Vertical slices through the $x=4.0$ plane for the WT1 simulation, showing the density perturbation (top) and contours of constant $B_y/B_0$ (middle) at $t=0.36$. The lower plot shows contours of constant $B_y/B_0$ at $t=0.80$. In the middle plot, the contours are evenly spaced in the range $-0.09 \le B_y/B_0 \le 0.85$. In the lower plot, the corresponding range is $-0.06 \le B_y/B_0 \le 0.55$}
\label{fig7}
\end{figure}

\begin{figure}
\resizebox{\hsize}{!}{\includegraphics{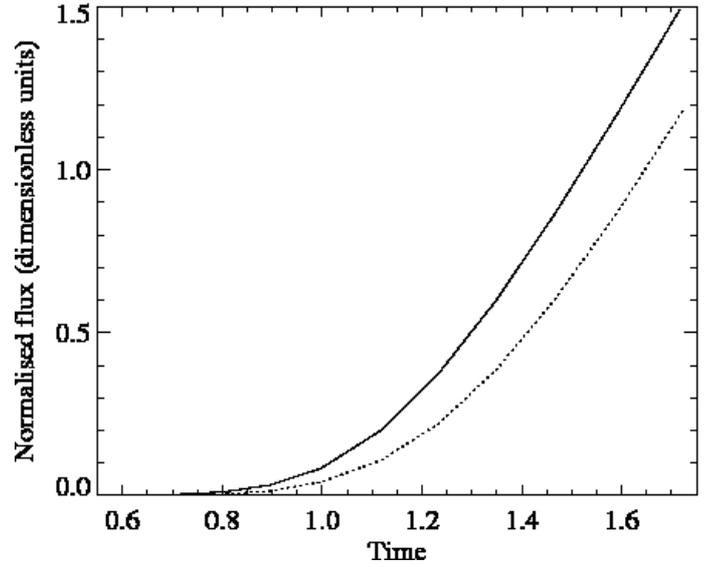}}
\caption{As figure~\ref{fig5}, but for simulations WT1 and L1. The solid line
  corresponds to the wide tube case (WT1), whilst the dotted line corresponds to
  the thin tube case (L1).} 
\label{fig8}
\end{figure}

We also considered variations in the twist and the radius of the
initial magnetic flux tube. Our aim here was to determine whether it
was possible to change the geometry of the tube in such a way as to
enhance the effects of magnetic buoyancy. Given that this was the
only lower entropy case in which there are any indications of magnetic
buoyancy, we used the strong field simulation L1 ($\beta=1.5$) as the
basis for investigating these variations in the flux tube geometry. In
this case, increasing the original twist by a factor of $10$ does not
seem to influence the evolution of the tube, or the flux emergence
rate. Although this enhancement of the twist does imply that the
fieldlines now make approximately 3-4 full turns around the axis of
the imposed flux tube, the resulting magnetic tension is still much
smaller than the initial magnetic pressure gradient across the radius of the tube. It is therefore not
surprising that the dynamics are largely unaffected by this change.  

\par Increasing the width of the flux tube has a much more significant
effect upon this system. Case WT1 is identical to L1 in every respect
except for the value of $R$, which has been increased from $0.1$ to
$0.15$. This more than doubles the initial horizontal magnetic flux in
the initial condition. The upper two plots in Figure~\ref{fig7} show
the density perturbation and the contours of $B_y/B_0$ in the $x=4.0$
plane at $t=0.36$. At this instant in time there is (as in case L1)
some indication of modest convective disruption, although there is
still a coherent flux tube with a pronounced density deficit. Indeed,
at $z=0.75$, the minimum density is approximately $60\%$ of the mean
density at this depth. So even after the initial phase of evolution,
this is still a configuration that should be susceptible to magnetic
buoyancy. The lower plot in Figure~\ref{fig7} shows the distribution
of $B_y/B_0$ in the same plane at $t=0.8$. This plot shows that some
coherent flux concentrations have risen into the upper layers of the
domain. Comparing this plot with the corresponding plot for L1 at
$t=0.80$ (which is shown in the top left part of Figure~\ref{fig4}),
we see that there seems to be more flux in the upper layers of the domain in the wider tube case. To illustrate this point in more quantitative terms, Figure~\ref{fig8} shows the (unsigned) flux emergence as a function of time for the WT1 simulation, plotted alongside the equivalent flux emergence curve for L1. In each case, the flux is normalised by the initial imposed flux in the $y$-direction. Given that the (normalised) flux emergence rate is significantly higher in case WT1, we can conclude that magnetic buoyancy is playing a much more significant role in the evolution of this wider magnetic flux tube. This suggests that the efficiency of magnetic buoyancy relative to the effects of convective disruption is determined not only by the initial peak magnetic field strength, but also by the width of the imposed flux tube.

\begin{figure*}
\centering
\includegraphics[width=10cm]{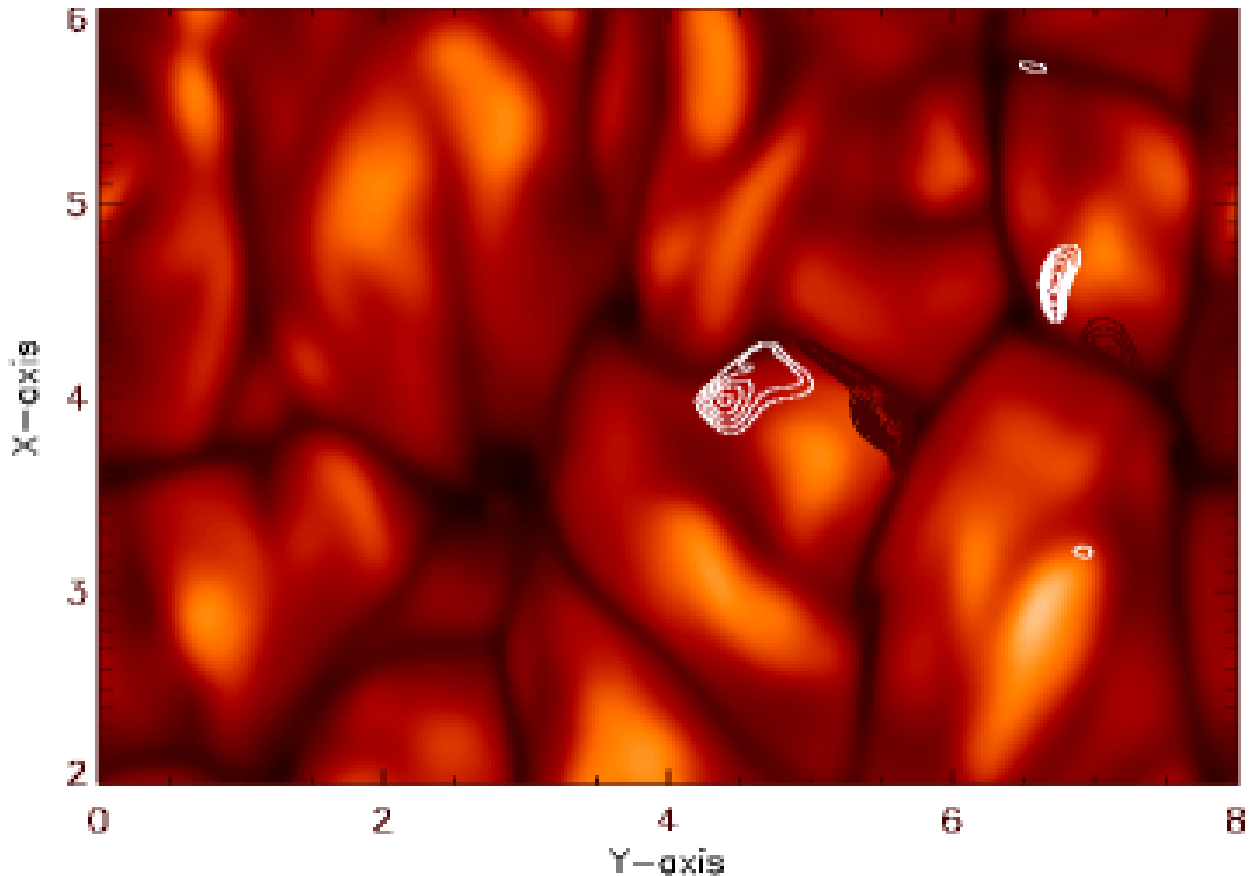}
\includegraphics[width=8cm]{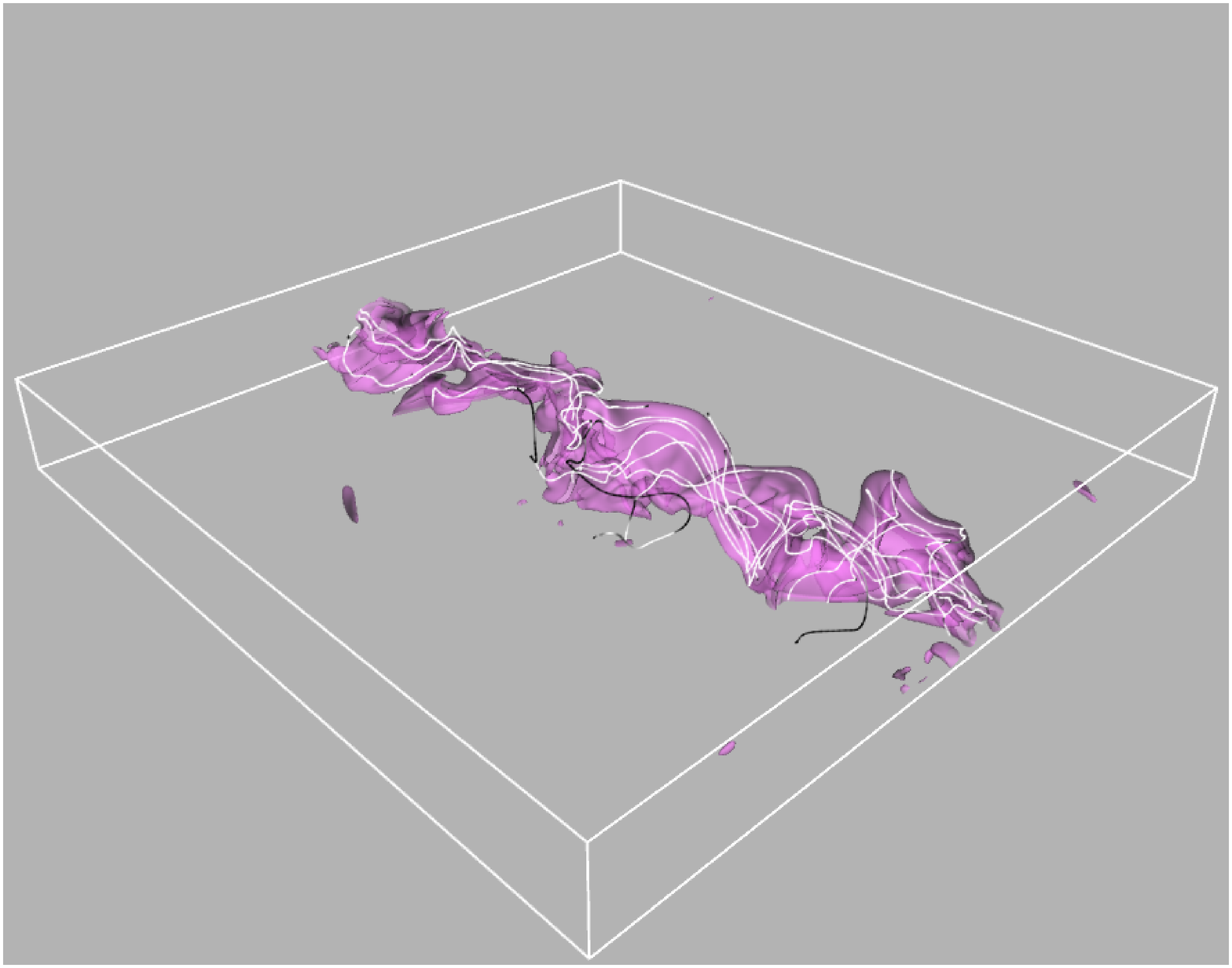}
\includegraphics[width=10cm]{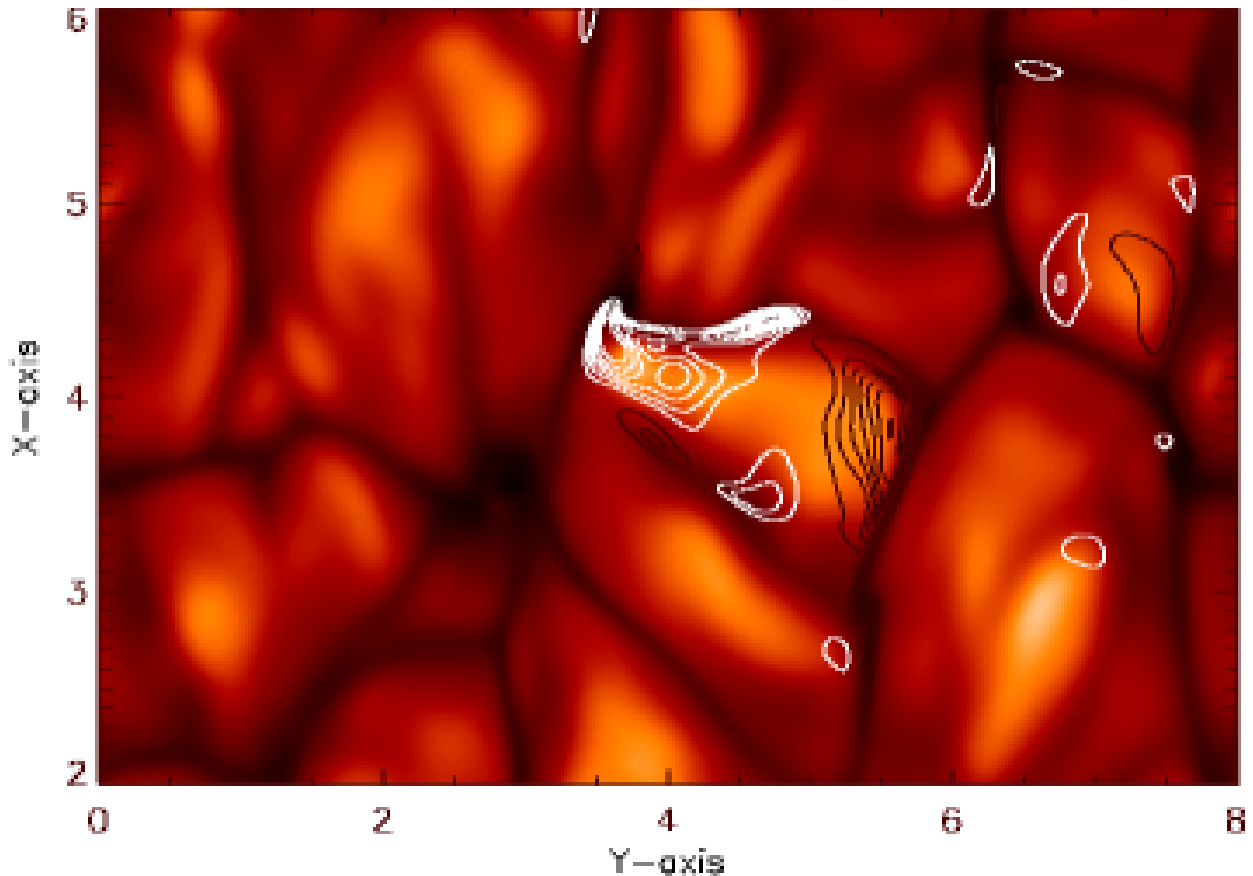}
\includegraphics[width=8cm]{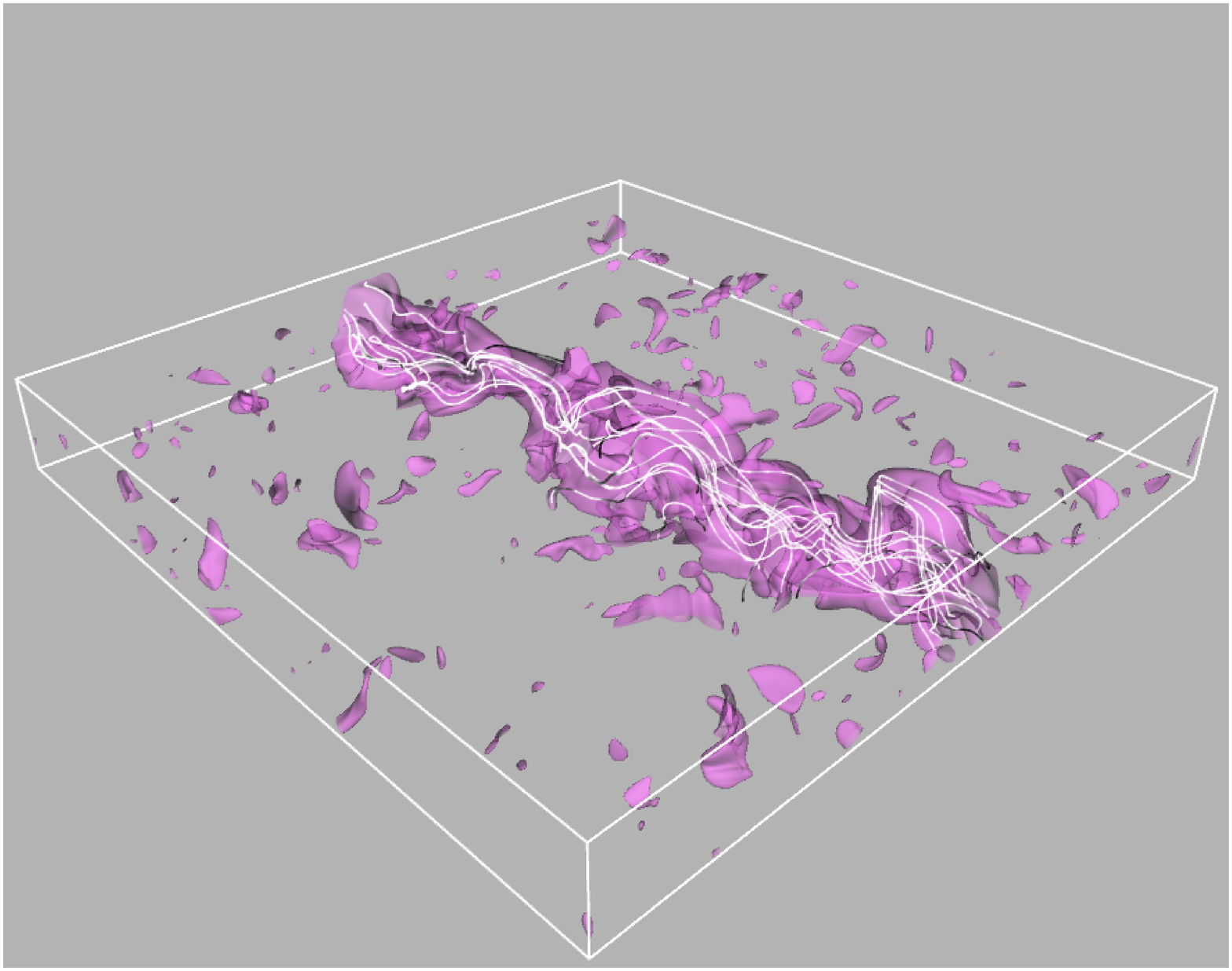}
\caption{Simulations L1 (top) and WT1 (bottom) at $t=1.59$. The plots on the left-hand side show the temperature distribution in a horizontal plane just below the upper surface of the computational domain (shaded contours) and the distribution of $B_z$ in the same horizontal plane (black and white contours). The volume renderings on the right-hand side of this figure show isosurfaces of $|\mathbf{B}|=0.25$. Some fieldlines have also been traced out along these isosurfaces in order to aid visualisation.}
\label{fig9}
\end{figure*}

\par The left-hand side of Figure~\ref{fig9} shows the flux emergence patterns for cases L1 and WT1 at $t=1.59$. In the L1 case, the vertical magnetic field distribution in the near-surface region is dominated by two well-defined bipolar magnetic regions. As illustrated in Figure~\ref{fig9}, these flux concentrations emerge initially within the warm granular interiors, in regions of convective upflow. However once the magnetic fields reach the near-surface layers, the horizontal convective motions start to sweep these flux concentrations into the intergranular lanes. So at later times these emerging vertical flux concentrations accumulate in the convective downflows. The flux emergence pattern in case WT1 is similar, although more flux has emerged, and the resulting magnetic field distribution is much less localised than the flux emergence pattern from case L1. The volume renderings on the right-hand side of Figure~\ref{fig9} show isosurfaces of the magnetic field strength ($|\mathbf{B}|=0.25$). Although the flux tube is wider in the WT1 case, the flux tube structure is similar in both cases. It is clear that the emerging bipoles that are observed at the surface belong to the same undulating flux tube, and are therefore connected by subsurface magnetic field lines. So in each case, the flux tube has resisted convective disruption effectively enough to maintain some semblance of its initial structure throughout the flux emergence process. 

\subsection{Varying the magnetic Reynolds number}

All of the calculations that have been described so far have a magnetic Reynolds number of $\mathcal{R}m\approx 140$. Although the ohmic decay time based on the depth of the layer, $\tau_{\eta}$, is approximately 140 convective turnover times for $\mathcal{R}m\approx 140$, the decay time based upon the width of the flux tube is approximately two orders of magnitude smaller than that. Hence, magnetic dissipation must be playing some role in the evolution of the system on the timescales of interest. The crucial question is whether or not the evolution of the system is {\it dominated} by magnetic dissipation. To investigate this issue, we varied the value of $\mathcal{R}m$. Cases L4, L5 and L6 are identical to L1, L2 and L3 in every respect except for the fact that $\mathcal{R}m\approx 70$, as opposed to $\mathcal{R}m\approx 140$. In cases L7, L8 and L9, we adopted a higher value of  $\mathcal{R}m\approx 280$. We also carried out some wider tube calculations (WT2 and WT3) with these values for the magnetic Reynolds number. As described in the Section 2.2, these choices for $\mathcal{R}m$ ensure that $Pm<1$ in all cases, and this restriction upon $Pm$ is the main reason why higher values of $\mathcal{R}m$ were not considered in the present study.  

\begin{figure}
\resizebox{\hsize}{!}{\includegraphics{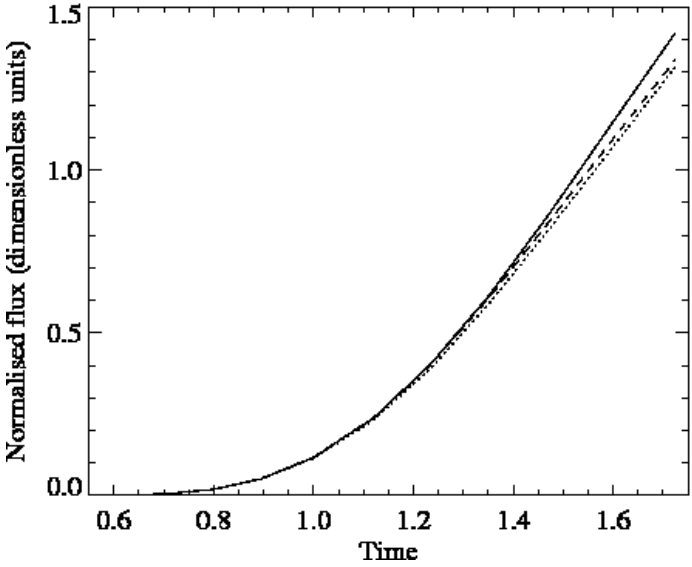}}
\caption{As figure~\ref{fig5}, but this time for the $\mathcal{R}m\approx 70$ thin tube cases (L4, L5 and L6). The solid line
  corresponds to L4 ($\beta=1.5$), the dotted line corresponds to
  L5 ($\beta=5$), whilst the dashed line corresponds to L6 ($\beta=35$).}
\label{fig10}
\end{figure}

\begin{figure}
\resizebox{\hsize}{!}{\includegraphics{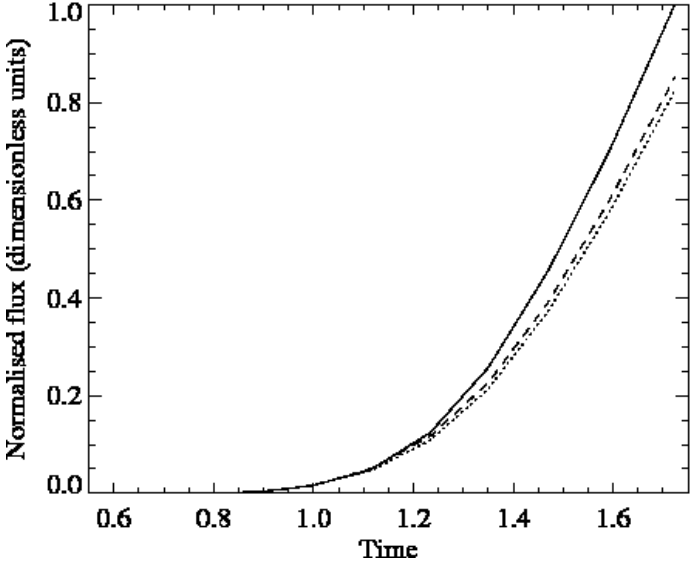}}
\caption{As figure~\ref{fig5}, but this time for the $\mathcal{R}m\approx 280$ thin tube cases (L7, L8 and L9). The solid line
  corresponds to L7 ($\beta=1.5$), the dotted line corresponds to
  L8 ($\beta=5$), whilst the dashed line corresponds to L9 ($\beta=35$).}
\label{fig11}
\end{figure}

\par Figures~\ref{fig10} and~\ref{fig11} show the normalised flux emergence rates for the thin flux tube, lower entropy cases with $\mathcal{R}m\approx 70$ and $\mathcal{R}m\approx 280$ respectively. These plots should be compared with Figure~\ref{fig5}, which shows the corresponding plot for the $\mathcal{R}m\approx 140$ cases. The first point to notice is that that $\beta$ dependence of the flux emergence rate is strongly-dependent upon $\mathcal{R}m$.  In particular, the enhancement of the flux emergence rate due to magnetic buoyancy in the $\beta=1.5$ simulations becomes much more pronounced at higher values of the magnetic Reynolds number. After one convective turnover time (at $t\approx 1.55$), the normalised surface fluxes in the $\mathcal{R}m\approx 70$ simulations differ by only a few percent whereas in the $\mathcal{R}m\approx 280$ simulations the normalised fluxes differ by approximately $15\%$ after a similar time period. We can conclude from this that magnetic buoyancy is more efficient at higher magnetic Reynolds number (presumably a consequence of the fact that there is less magnetic dissipation), provided that the magnetic field is strong enough initially to produce a magnetic pressure that is a significant fraction of the local gas pressure. However it is not just the efficiency of magnetic buoyancy that is dependent upon the magnetic Reynolds number. Comparing Figures~\ref{fig10} and~\ref{fig11}, we see that the flux emergence rates in the higher magnetic Reynolds number cases are systematically {\it lower} than the corresponding flux emergence rates in the smaller $\mathcal{R}m$ calculations. Although not shown here, the same is true in the wider tube calculations. The reason for this reduced flux emergence rate is that convective disruption (like magnetic buoyancy) is also more efficient at higher magnetic Reynolds numbers, because there is a greater tendency for magnetic field lines to be advected by the flow. As a result of this, loops of flux are continually removed from the tube, before being reprocessed by the local convective motions. This causes less flux to emerge at the surface of the domain. As described in the Introduction, we know from previous studies \citep[][]{DORCH01} that convective motions generally reduce the efficiency of flux emergence. We can now make the stronger statement that this reduction in the flux emergence rate due to convective disruption is more pronounced at higher magnetic Reynolds numbers in the $Pm<1$ parameter regime. 

\par
\begin{figure*}
\centering
\includegraphics[width=10cm]{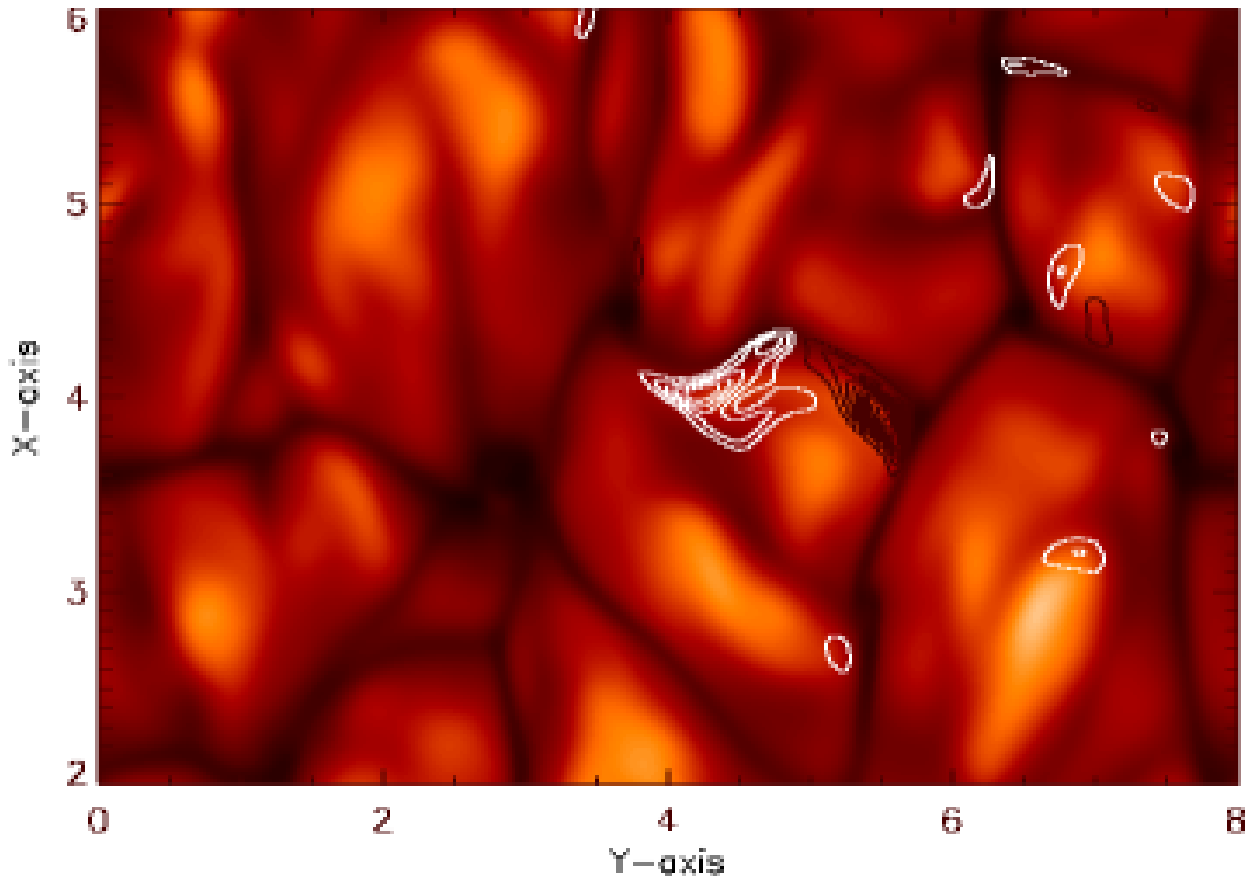}
\includegraphics[width=8cm]{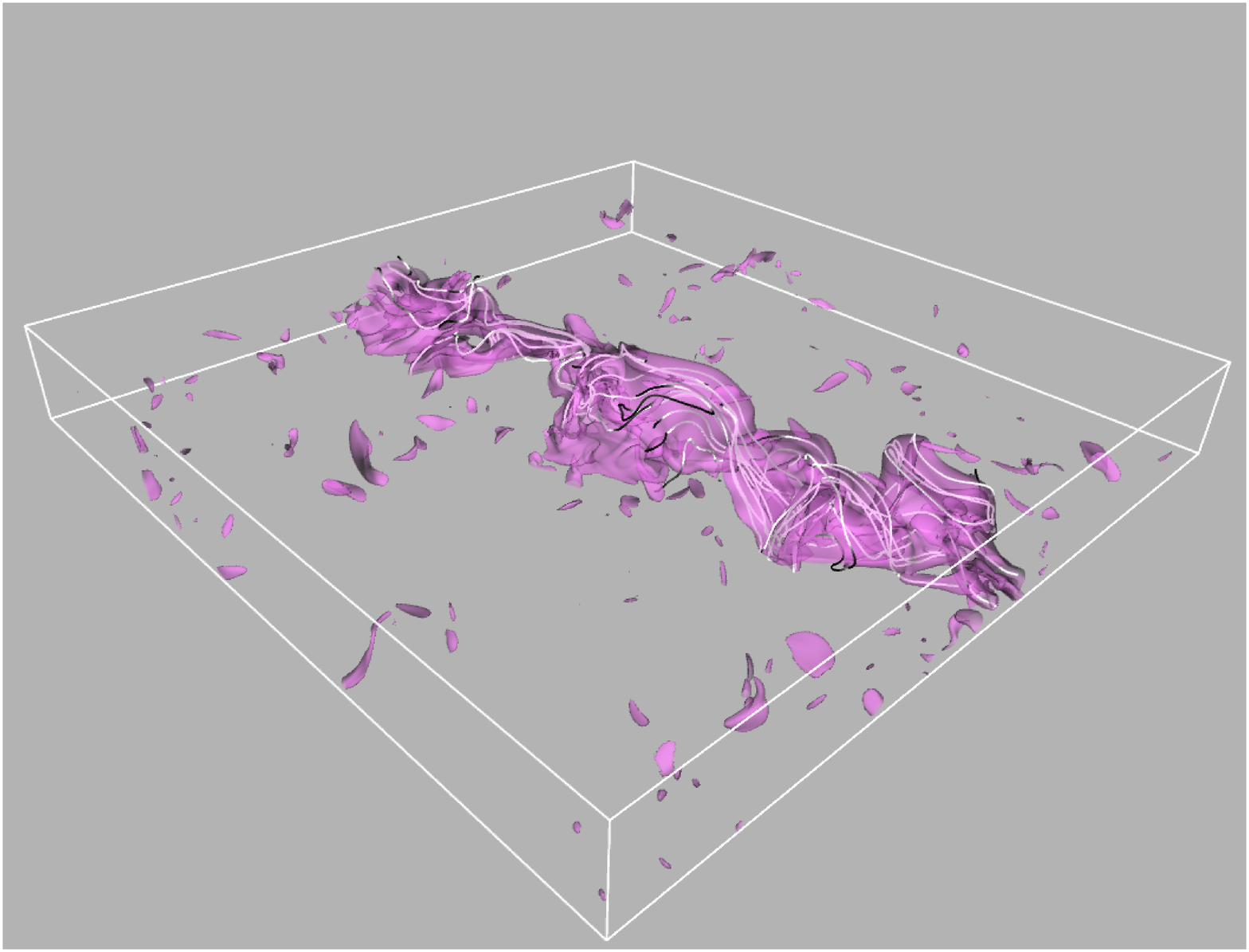}
\includegraphics[width=10cm]{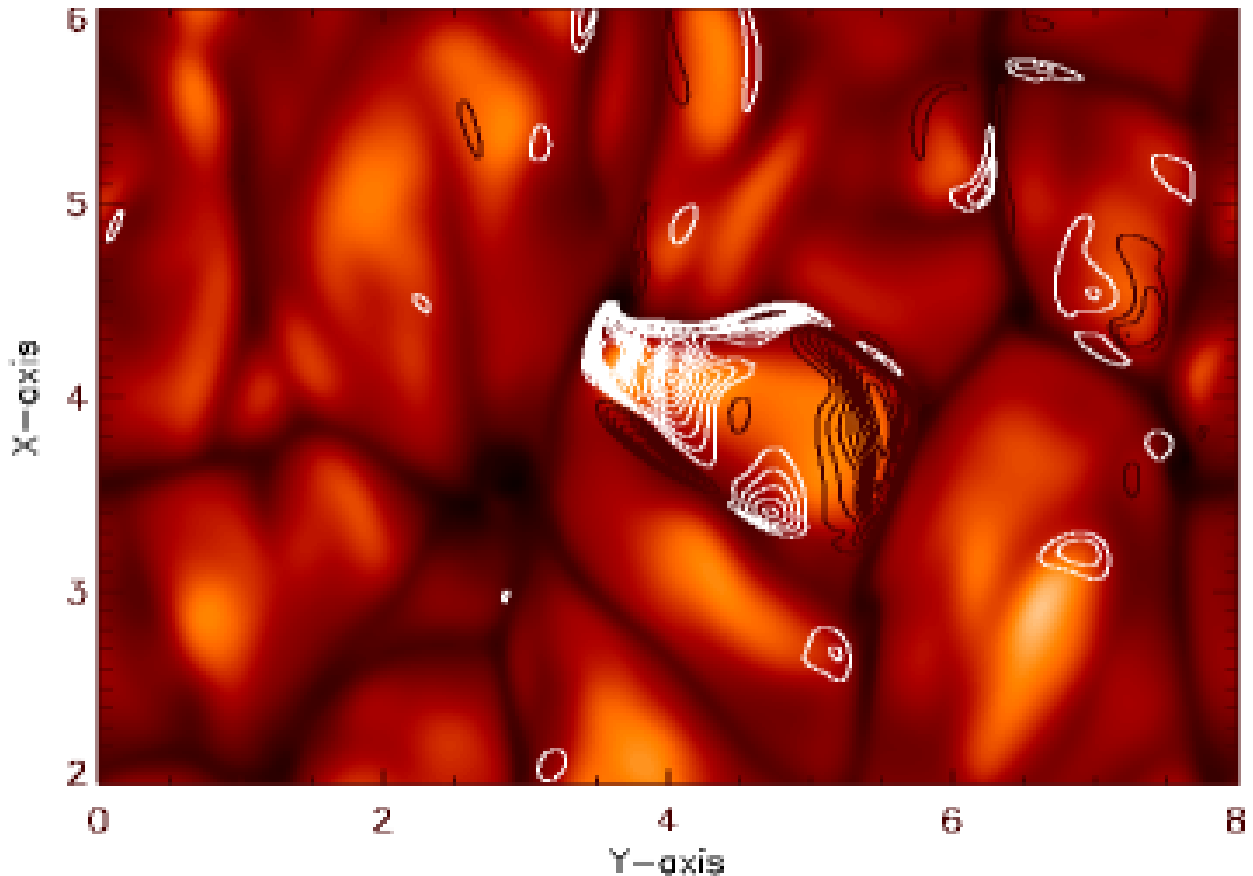}
\includegraphics[width=8cm]{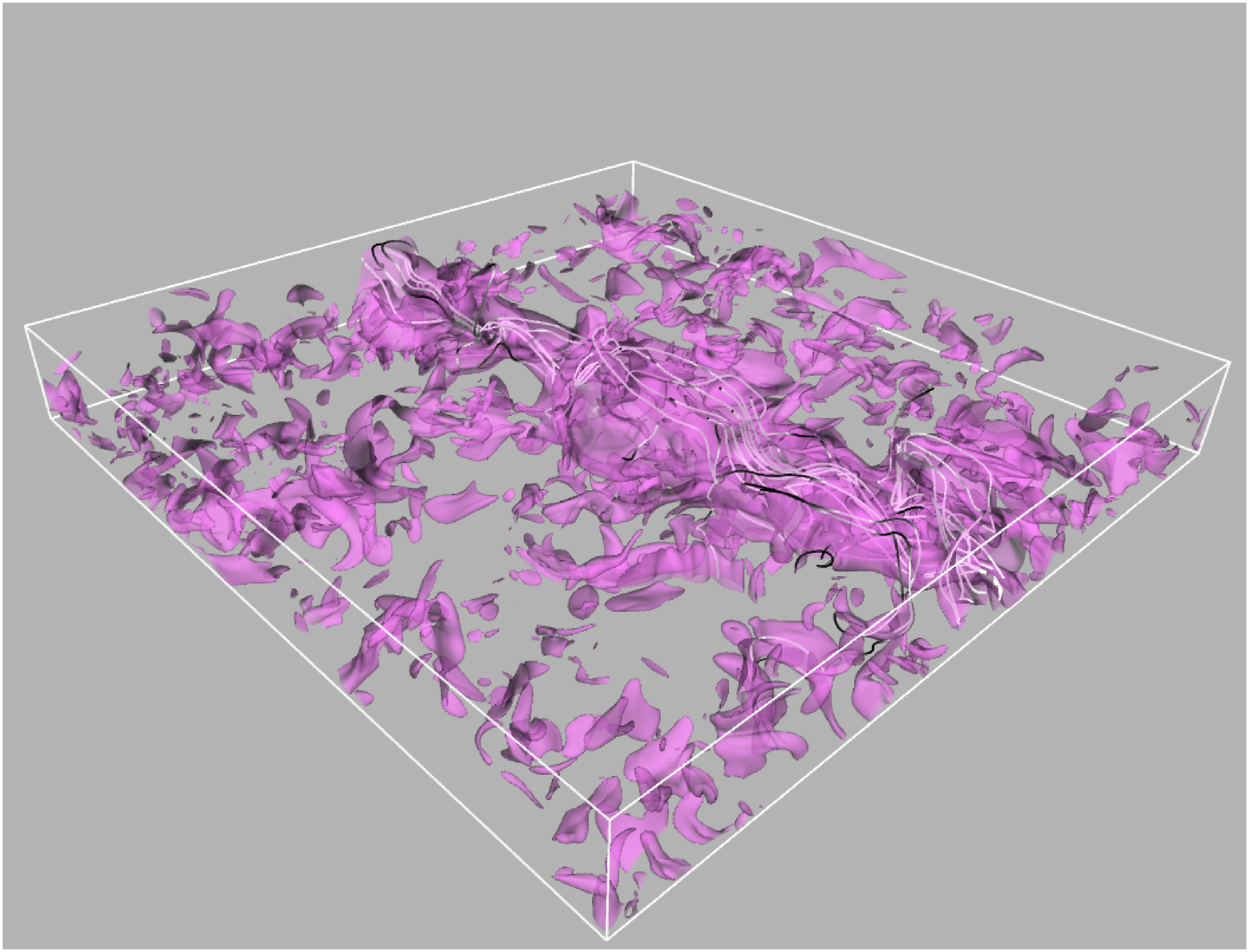}
\caption{As figure~\ref{fig9}, but here for the corresponding $\mathcal{R}m\approx 280$, lower entropy cases, L7 (top) and WT3 (bottom).}
\label{fig12}
\end{figure*}

 Given that the flux emergence rates are strongly dependent upon $\mathcal{R}m$, we would also expect the flux emergence patterns to show a similar $\mathcal{R}m$-dependence. Figure~\ref{fig12} shows the flux emergence patterns at $t=1.59$, alongside a volume rendering of the associated magnetic flux tubes, for cases L7 and WT3 (the lowest $\beta$, lower entropy cases for $\mathcal{R}m\approx 280$). Although the strongest emerging flux concentrations are still found in the vicinity of $x=4.0$, which corresponds to the location of the axis of the initial magnetic flux tube, many other weaker magnetic flux concentrations are also observed. This indicates that the flux tube has undergone more convective shredding in these higher magnetic Reynolds number cases. This convective disruption is highlighted even more clearly in the volume renderings, particularly in the wide tube case where the magnetic field distribution is highly disordered and fragmented. Although not shown here, very little convective shredding is observed in the $\mathcal{R}m\approx 70$ cases, and the flux emergence pattern is dominated by a small number of coherent dipoles. 

\subsection{Non-equilibrium expanding flux tubes}

In this subsection, we consider the last two cases listed in
Table~\ref{table2}, namely E1 and E2. Here, we make no initial perturbation
to the flow, pressure or entropy at $t=0$. The idea of these simulations is
to allow the convection to select its own ``initial condition'', by
adjusting itself in response to the presence of a dynamically-significant magnetic field. Any
magnetic buoyancy is therefore induced in a self-consistent way,
albeit from a non-equilibrium initial condition. This approach, whilst
clearly oversimplified (and unrealistic for the solar convection zone), has the great advantage that it is not necessary
to specify a particular distribution for any of the thermodynamic quantities along the tube, in
contrast to the other cases that are considered in this paper. Given
that there is considerable uncertainty regarding the most suitable initial
conditions for idealised calculations of this type, we would argue
that simulations E1 and E2 are useful illustrative calculations in this context.

\par Obviously, the initial phase of evolution of these simulations is
dominated by the effects of the Lorentz force. The radial magnetic
pressure gradient tends to drive fluid away from the axis of the
tube. Furthermore, this occurs on an Alfv\'enic timescale, which means
that this process occurs very rapidly (in these cases, the Alfv\'en speed at the depth of the flux tube is approximately 4 times larger than a typical convective velocity). This pressure-driven radial outflow 
partially evacuates the flux tube. It also causes the flux tube to
expand in the radial direction. By conservation of flux, this
expansion leads to a reduction in the mean magnetic field strength
within the magnetic flux tube. As a result of this expansion, the flux tube
rapidly reaches a state in which it is in (near) total pressure
balance with its surroundings. Given that this adjustment is very
rapid, convective perturbations play a negligible role during this
early phase of evolution. 

\par
\begin{figure}
\resizebox{\hsize}{!}{\includegraphics{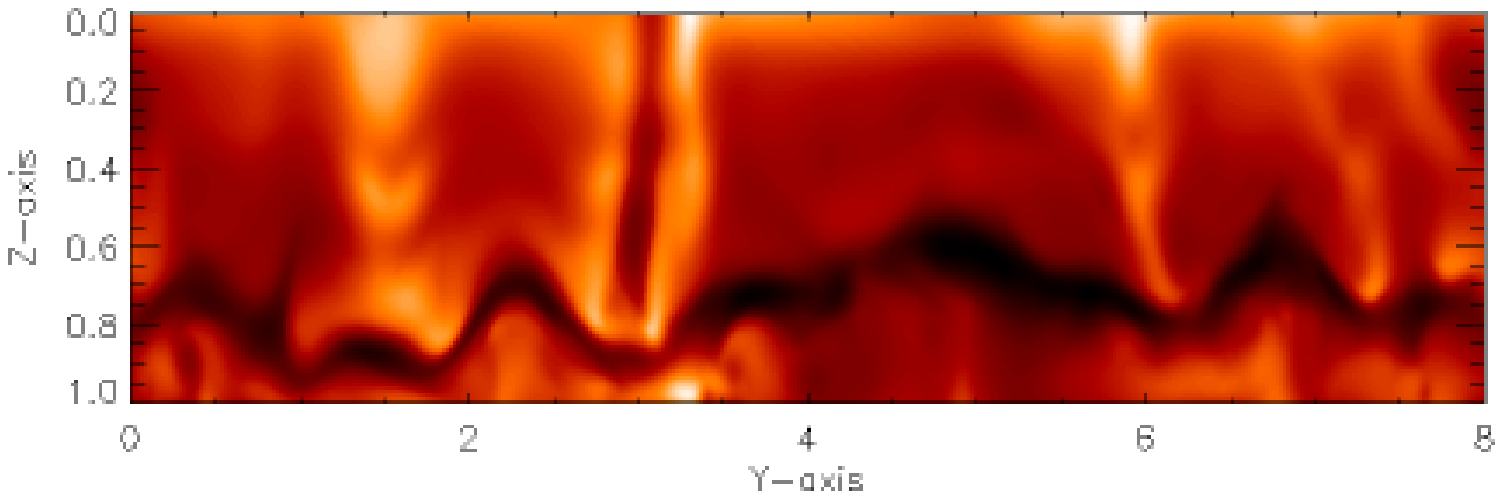}}
\resizebox{\hsize}{!}{\includegraphics{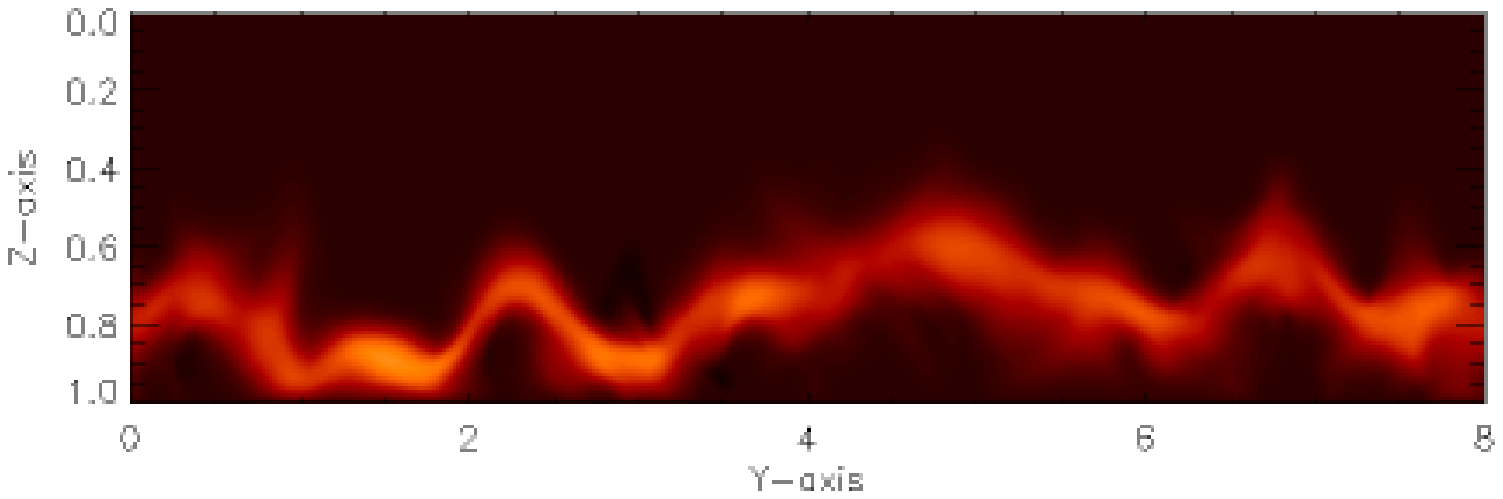}}
\resizebox{\hsize}{!}{\includegraphics{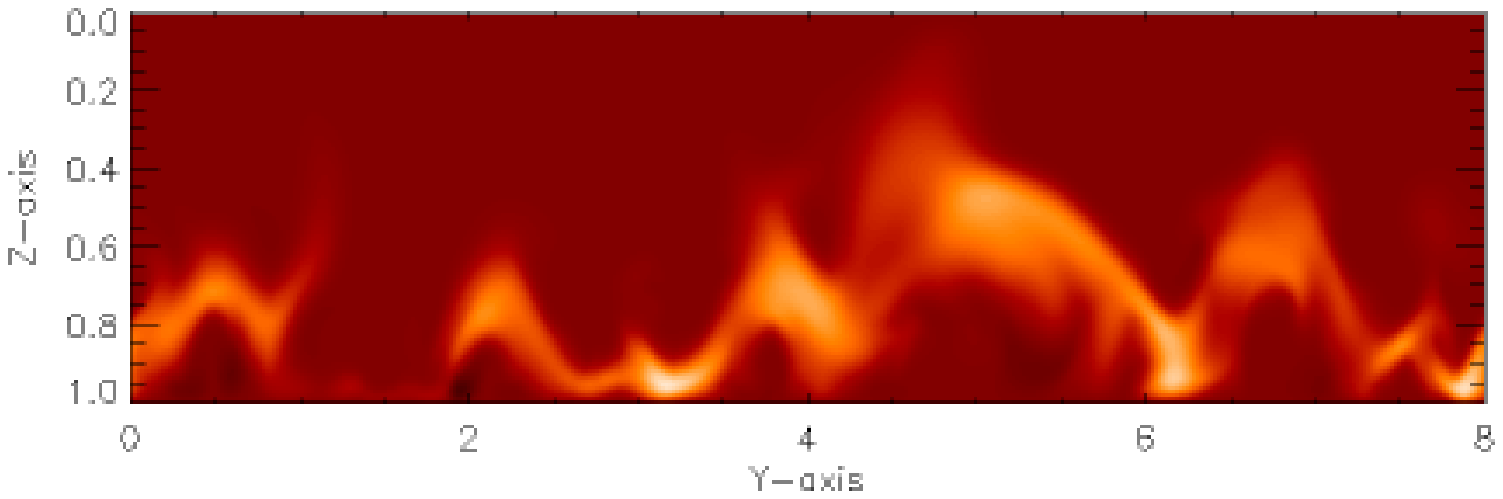}}
\caption{Vertical slices through the $x=4.0$ plane for the E1 simulation, showing the density perturbation (top) and contours of constant $B_y/B_0$ (middle) at $t=0.36$. The lower plot shows contours of constant $B_y/B_0$ at $t=0.80$. In the middle plot, the contours are evenly spaced in the range $-0.13 \le B_y/B_0 \le 0.9$. In the lower plot, the corresponding range is $-0.25 \le B_y/B_0 \le 0.45$.}
\label{fig13}
\end{figure}

\begin{figure}
\resizebox{\hsize}{!}{\includegraphics{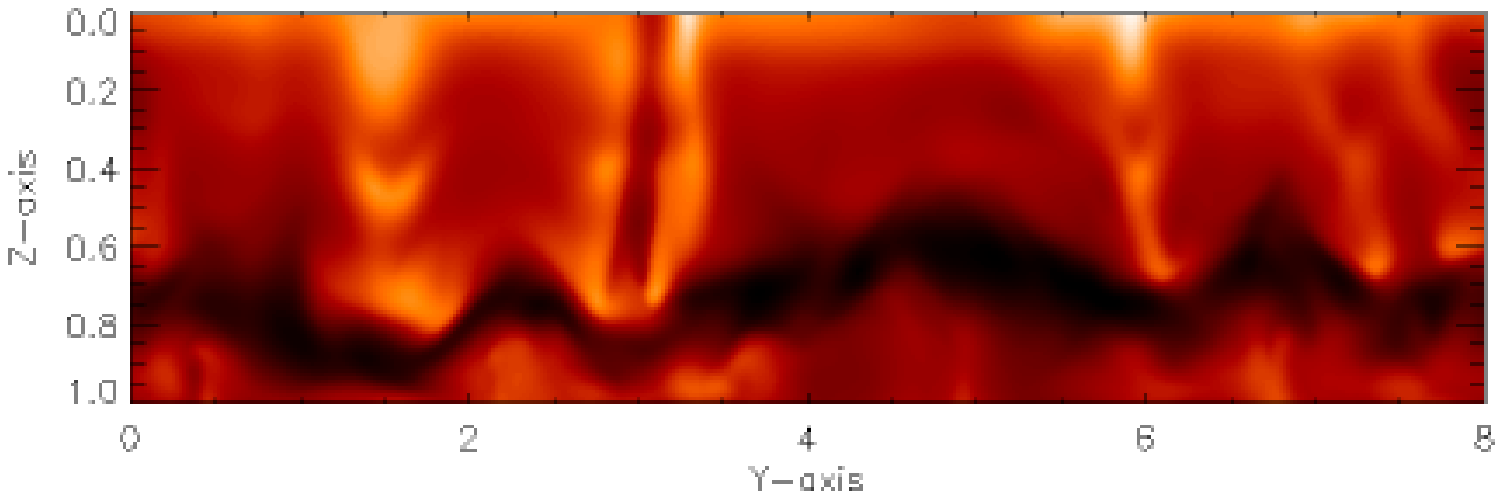}}
\resizebox{\hsize}{!}{\includegraphics{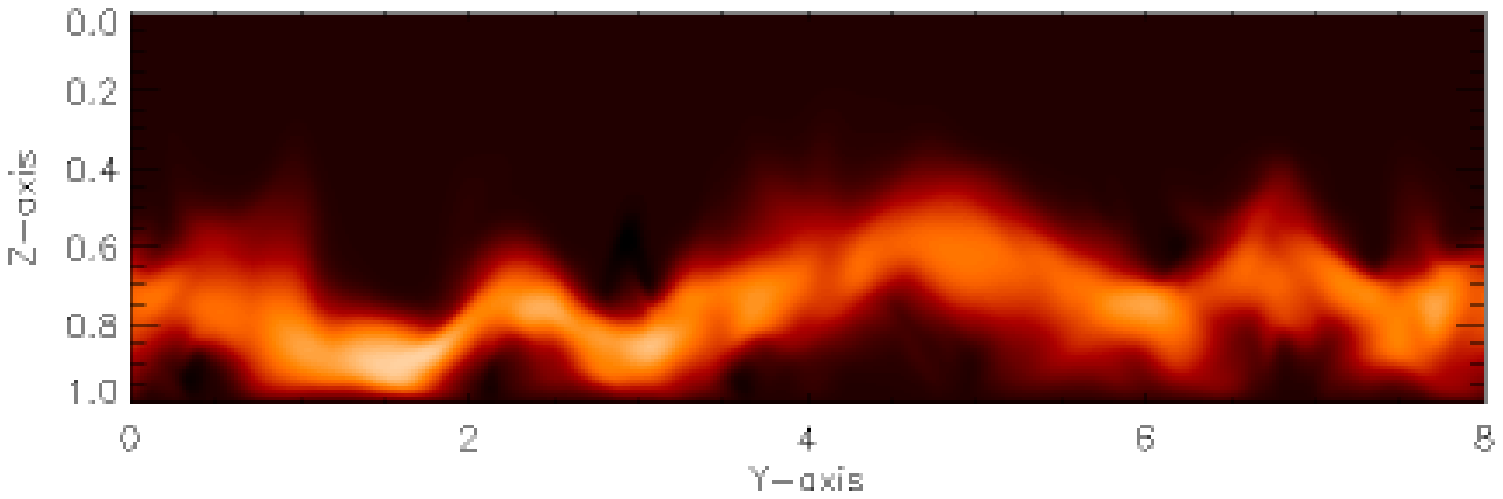}}
\resizebox{\hsize}{!}{\includegraphics{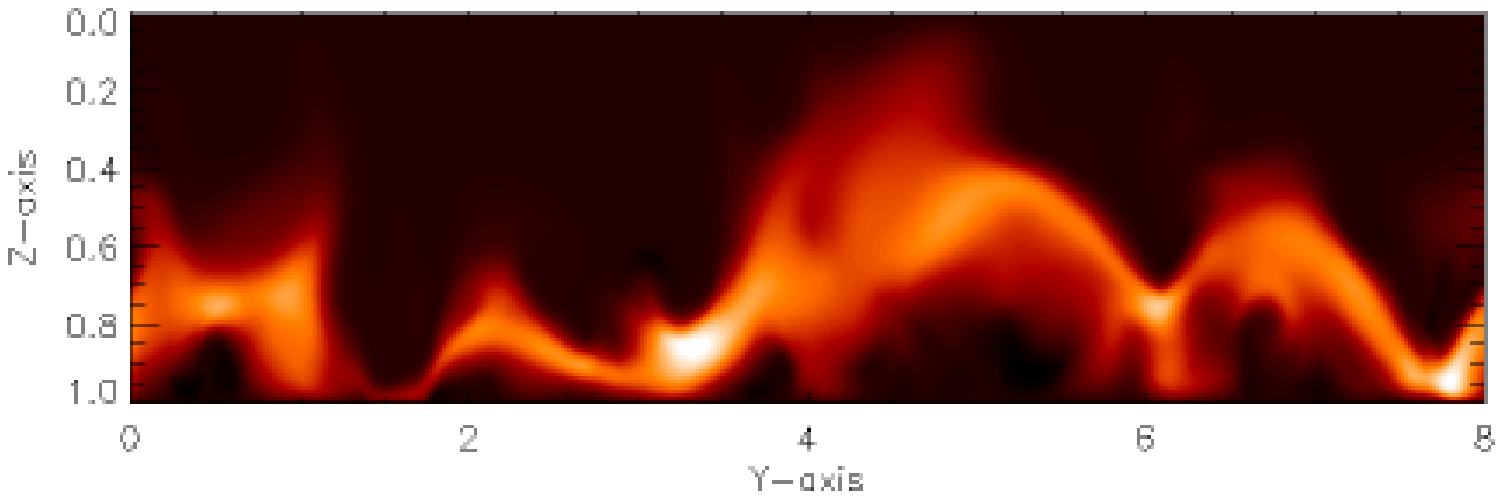}}
\caption{Like Figure~\ref{fig13}, but this time for the E2 simulation. In the middle plot, the contours are evenly spaced in the range $-0.09 \le B_y/B_0 \le 0.85$. In the lower plot, the corresponding range is $-0.06 \le B_y/B_0 \le 0.55$.}
\label{fig14}
\end{figure}

\begin{figure}
\resizebox{\hsize}{!}{\includegraphics{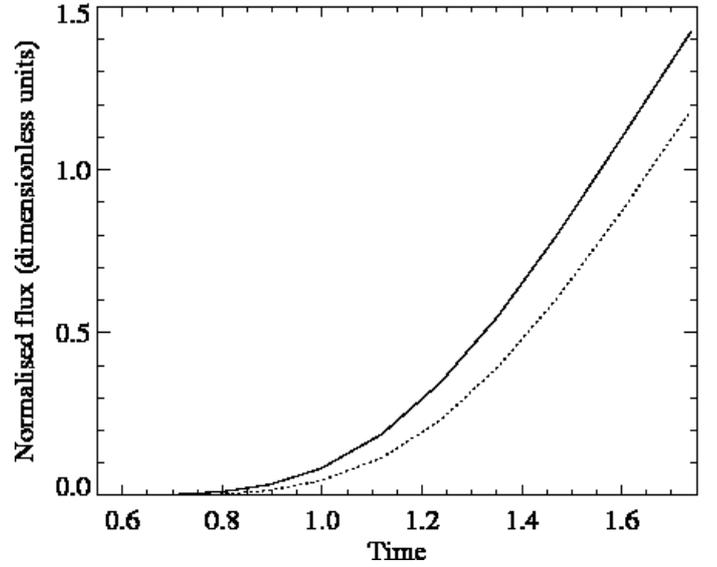}}
\caption{As figure~\ref{fig5}, but for simulations E1 and E2. The solid line
  corresponds to the wide tube case (E2), whilst the dotted line corresponds to
  the thin tube case (E1).} 
\label{fig15}
\end{figure}

 Having established that these cases rapidly evolve towards
pressure-balanced states, we now consider the subsequent evolution of
each of these calculations. The initial magnetic field in case E1 is
identical to that of case L1, with $\beta=1.5$ and $R=0.1$. The top
two plots in Figure~\ref{fig13} show a snapshot of this case at
$t=0.36$. The upper plot shows the density perturbation in the $x=4.0$
plane, whilst the middle plot shows the $B_y$ distribution in the same
plane (normalised by $B_0$). At this instant in time, the minimum
density within the flux tube is approximately $75\%$ of the mean
density at $z=0.75$. Although this level of partial evacuation is not
negligible, it is modest compared to some of the other cases. The
$B_y/B_0$ distribution at $t=0.36$ is comparable to that shown in the
upper plot of Figure~\ref{fig3} (where the same contour spacings were
used). The lower plot in Figure~\ref{fig13} shows the $B_y/B_0$
distribution in the $x=4.0$ plane at $t=0.8$. Again this horizontal
magnetic field distribution is comparable to that of the corresponding
plot for the L1 case (shown in the top left plot in
Figure~\ref{fig4}). In qualitative terms, these plots clearly indicate
that the flux tube in this new calculation has evolved in a rather
similar way to the flux tube from the L1 case, despite the unrealistic
initial conditions. Figure~\ref{fig14} shows the corresponding results
for the E2 case ($\beta=1.5$ and $R=0.15$), which can be compared
directly to WT1 (as illustrated in Figure~\ref{fig7}). At $t=0.36$,
the minimum density at $z=0.75$ for this E2 case is approximately
$70\%$ of the mean density at this depth, which is slightly higher
than the minimum density at this level at the corresponding stage of
the WT1 simulation. However, in almost every other respect, the
results from this E2 case are qualitatively rather similar to those of
the WT1 simulation. The flux emergence rates for cases E1 and E2 are
shown in Figure~\ref{fig15}. These are clearly comparable to the
corresponding flux emergence curves for the L1 and WT1 cases. So
although the initial conditions in cases E1 and E2 are rather
idealised (and probably unrealistic), we have shown that these
simulations do evolve in a similar way to our previous
calculations. This suggests that local simulations of this type are
relatively insensitive to the precise choice of initial conditions,
with the evolution of the system (for a given value of $\mathcal{R}m$)
being determined primarily by the initial magnetic field
distribution. Given the wide range of plausible choices of initial
conditions in local models of this type, this is an encouraging result.

\section{Summary and discussion}

We have used numerical simulations to investigate the dynamical 
evolution of a magnetic flux tube embedded within a granular convective layer. Throughout the calculations we have adopted the same basic hydrodynamic flow, with a Reynolds number of $\mathcal{R}e \approx 420$. When the magnetic flux tube is introduced, the gas pressure is adjusted so that the tube is in pressure balance with its surroundings. The tube is then made buoyant by choosing an appropriate entropy distribution. The subsequent evolution of this tube is influenced by two key processes, namely convective disruption and magnetic buoyancy. However the extent to which magnetic buoyancy plays a role in the flux emergence process depends in a rather subtle way upon the model parameters. In this systematic survey, we varied the initial peak magnetic field strength, the twist and width of the flux tube, the magnitude of the entropy along the tube axis and the magnetic Reynolds number. In all cases, the magnetic Prandtl number, $Pm=\mathcal{R}m/\mathcal{R}e$ is less than unity, as would be expected for the solar convection zone. This is the first study of this type to explore this low $Pm$ regime. 

\par At a moderate magnetic Reynolds number of $\mathcal{R}m\approx
140$, the evolution of a thin magnetic flux tube is partially
determined by the effects of convective disruption. In the weak field
(high $\beta$) cases, the magnetic field appears to play a relatively
passive role in the dynamics. However, there is an enhanced flux
emergence rate in the strongest field cases due to the effects of
magnetic buoyancy. Increasing the magnitude of the initial peak
entropy along the tube axis does not change the $\beta$ dependence,
but it does increase the rate of flux emergence in all cases. Varying
the twist of the flux tube does not seem to influence the evolution of
the system (presumably because the magnetic tension is still much
weaker than the magnetic pressure gradients in the flux
tube). However, increasing the width of the tube does produce an
enhanced flux emergence rate in the strong field case. This is
presumably because a thicker tube is more robust, and tends to resist
convective disruption more effectively than a thinner flux tube. As a result of this, the flux tube maintains its coherence more effectively during the early stages of evolution, which allows magnetic buoyancy to operate in a more efficient manner. So our main conclusion from the $\mathcal{R}m\approx 140$ cases is that the contribution of magnetic buoyancy to the flux emergence process is directly related not only to the peak field strength, but also to the width of the initial magnetic flux tube. 
\par We also varied the magnetic Reynolds number in this system in
order to determine the extent to which the evolution of the tube is
influenced by the effects of magnetic dissipation. This parametric
survey produced two key results. Firstly, magnetic buoyancy
contributes most effectively to the flux emergence process in the low
$\beta$, high magnetic Reynolds number regime. Secondly, for a given value of $\beta$, the flux
emergence rate is always lower in the higher $\mathcal{R}m$ regime. This is due to the fact that
convective disruption is much more efficient at higher magnetic
Reynolds number (where field lines are more easily advected by the flow). The extent to which this result depends upon the magnetic Prandtl number is unclear, although moving to higher $\mathcal{R}m$ (thereby increasing $Pm$) would be computationally simpler than reducing $Pm$ by increasing $\mathcal{R}e$. At higher $\mathcal{R}m$ this convective flow should be capable of sustaining a small-scale dynamo, although it is not clear how such a flow would interact with an emerging flux tube. This is a possible area for future work.  

\par Motivated by some of the uncertainty surrounding the most
appropriate choice of initial conditions for problems of this type, we
also carried out some idealised simulations in which the gas pressure
was not adjusted when the flux tube was introduced. We stress that
this is not a realistic situation for the solar convection zone given
that this leads to a strong imbalance between the magnetic pressure
and the local gas pressure. However, this does allow the system to
adjust to the presence of the flux tube in a self-consistent way. This
adjustment occurs very rapidly and (somewhat remarkably) the
subsequent evolution of the system is comparable to that of the
previous set of simulations, at least at comparable magnetic Reynolds
numbers. Results from these idealised calculations strongly suggest
that the only aspect of the initial conditions that influences the
evolution of the flux tube is the initial magnetic field
distribution. This is an encouraging result because it suggests that
flux emergence simulations are largely insensitive to the precise
choice of initial conditions, which is a matter of considerable
uncertainty in idealised models of this type.  

\bibliographystyle{aa}
\bibliography{paper}
\end{document}